\documentclass[11pt]{article}
\usepackage[
    top=2cm,
    bottom=2cm,
    left=3cm,
    right=3cm
]{geometry}
\usepackage{authblk}  
\usepackage{placeins}
\usepackage{amsmath}
\usepackage{graphicx}
\usepackage{multirow}
\usepackage{caption}
\usepackage{subcaption}
\usepackage[colorlinks=true, allcolors=blue, pdfencoding=auto]{hyperref}
\usepackage{cleveref}
\usepackage{tabularx}
\usepackage{ltablex}
\keepXColumns
\usepackage{csvsimple}
\usepackage{sidecap}
\usepackage{svg}
\usepackage[version=4]{mhchem}
\usepackage{siunitx}
\usepackage{booktabs} 
\usepackage{pdflscape}
\usepackage{float}
\usepackage{tablefootnote}

\usepackage[sorting=none, style=numeric, maxnames=3, minnames=1]{biblatex}
\title{VNS Tokamak for Medical Isotope Production\thanks{This work has been submitted to the Nuclear Science and Engineering PHYSOR 2026 Special Issue}}
    
\author[1,2]{Christopher Ehrich}
\author[3]{Pavel Pereslavtsev}
\author[3]{Christian Bachmann}
\author[1,2]{Christian Reiter}

\affil[1]{Forschungs-Neutronenquelle Heinz Maier-Leibnitz (FRM II), Lichtenbergstraße 1, Garching bei München, Germany}
\affil[2]{Chair of Applied Nuclear Technologies, School of Engineering and Design, Technical University of Munich, Boltzmannstraße 15, Garching bei München, Germany}
\affil[3]{EUROfusion Consortium, FTD Department, Boltzmannstraße 2, Garching bei München, Germany}

\begin{document}
\maketitle
\begin{abstract}
The Volumetric Neutron Source (VNS) tokamak is a proposed fusion reactor for testing components under fusion neutron irradiation, and has potential use for radioisotope production. The VNS geometry is modeled in the Serpent 2.2.2 and OpenMC 0.15.2 neutronics codes. Coupled neutron-photon simulations compared fluxes, spectra, and selected reaction rates in the blanket and vacuum vessel. Good agreement was found overall, with the largest difference found in ($n,2n$) reactions. On an HPC cluster, Serpent 2 was found to have shorter computation time in coupled simulations, while OpenMC was faster in neutron only simulations. Radioisotope production yields were simulated in Serpent 2.2.2 for capsule and Cobalt plate irradiation facilities. Results indicate potential for large volume production of \ce{^{99}Mo}, \ce{^{131}I}, \ce{^{225}Ac}, \ce{^{177}Lu}, \ce{^{192}Ir}, \ce{^{64}Cu}, \ce{^{67}Cu}, \ce{^{161}Tb}, and \ce{^{153}Sm} while \ce{^{203}Pb} indicates lower potential. \ce{^{100}Mo} and LEU target heating was calculated, suggesting the LEU target mass or the cooling may need adjustment. Optimized \ce{^{60}Co} production yielded \SI{1.2}{GBq\per\milli\gram} and 100,00\SI{0}{TBq} after a 3-year irradiation period. Sensitivity to plant outage for \ce{^{99}Mo}, \ce{^{131}I}, \ce{^{177}Lu}, and \ce{^{60}Co} was simulated, suggesting irradiation can be restarted for the same isotope loading and demonstrated long-lived \ce{^{60}Co} to be robust to long plant dwell-time.
\end{abstract}
\noindent\textbf{Keywords:} Fusion Neutronics, Medical Isotope Production, VNS Tokamak, Serpent, OpenMC
\section{Introduction}\label{sec:1}
Nuclear fusion has recently received significant political support in Germany, as it presents a potentially attractive technology for large scale clean energy production with an abundant fuel supply. The ``Aktionsplan Fusion'' \cite{BMFTR2025}, a measure of the Hightech Agenda Deutschland, has set a goal to build the world’s first fusion power plant in Germany. The most achievable fusion reaction is that of Deuterium and Tritium, D-T fusion, because it has the largest cross section and hence requires less extreme temperatures to occur. This reaction produces a high energy of 17.6 MeV. 14.1 MeV of the energy is carried away by a neutron, the remaining 20\% by a He nucleus. While the Deuterium can be provided from external sources, for future fusion energy applications, the Tritium supply must be produced inside the reactor by neutrons interacting with Lithium. For regular operation, e.g. in large commercial fusion power plants, Tritium breeding blankets are foreseen \cite{PEARSON20181140}. Tritium production in fission reactors is still required to provide the several kilograms of Tritium to startup a fusion power plant. The primary Tritium production reactions in fusion are:
\begin{align}
    ^6\ce{Li}+n &\rightarrow ^3\ce{T}+^4\ce{He}   +\text{\SI{4.8}{MeV}} \\
    ^7\ce{Li}+n &\rightarrow ^3\ce{T} +^4\ce{He}  +n -\text{\SI{2.5}{MeV}}
\end{align}
The $^7$\ce{Li} Tritium-production reaction can only take place above the \SI{2.5}{MeV} threshold energy, while $^6$\ce{Li} has a much larger thermal cross section, but negatively affects neutron balance. Thus the nature of the neutron economy of a fusion reactor requires a highly optimized Tritium breeding blanket, particularly for large scale fusion power reactors. 

\section{Volumetric Neutron Source (VNS)}

For the testing and qualification of fusion nuclear components and for the demonstration of the operation of a nuclear fusion plant including the fuel cycle, a volumetric neutron source (VNS) has been proposed \cite{Federici_2023}. The facility concept has been developed by EUROfusion and is mainly based on ITER technology \cite{BACHMANN2025114796, Bachmann_2026}. The plasma scenario has been assessed with state-of-the-art plasma simulation codes of different levels of fidelity up to integrated modeling \cite{sicciniophysics}. These have been benchmarked against tritium-rich JET plasma experiments \cite{Maslov_2023}. Neutrons are generated in the VNS plasma as a result of the injection of high-energy neutrals causing beam-target fusion of deuterium and tritium \cite{HOPF2025114870}. The neutron wall load is $\approx$\SI{0.5}{\mega\watt\per\meter\squared} \cite{BACHMANN2025114796} and consequent neutron flux is intense, ca. \SI{1e14}{\per\centi\meter\squared\per\second}  . Since the device is designed to be operated in continuous mode to realize the high neutron fluence expected in the core components of fusion power plants, it is, in principle, suitable also for the production of radioisotopes, which is addressed in this article.

An important aspect that differs VNS from fusion reactors targeting power generation is the far lower Tritium consumption ($<$\SI{1}{kg / a} \cite{BACHMANN2025114796, Bachmann_2026}) due to the much lower fusion power and small major radius. This could be provided for by current availability from Canadian and Korean CANDU reactors, with their combined fleets producing $\approx$\SI{2.7}{kg / a} \cite{PEARSON20181140,NI20132422}. Additionally, high-flux research reactors like FRM II, provide potential for a European Tritium supply chain. 

As Tritium breeding is foreseen in the VNS for testing purposes only, a large neutron surplus is available for additional nuclear reactions. One promising application of the VNS is radioisotope production. Thanks to the large volume, high flux, and absence of a direct feedback to the fusion reaction by neutron flux, magnetic confinement fusion reactors are excellent candidates for medical isotope production \cite{pereslavtsev2024potential}.

Because of the uniquely efficient $\ce{W}+\ce{TiH_2}$ shielding in the VNS geometry and the high leakage rate from additional external heating ports, there could be discrepancies in detector response estimates across codes, and it is crucial to have reliable simulation tools to assess the feasibility of the aforementioned applications in the VNS. If the reaction rates differ between codes, it follows that predictions for medical isotope production also can, and it is important to qualify the results. In this context, neutron flux, photon flux, ($n,T$), ($n,2n$) reaction rates, and the efficiency of high performance computer (HPC) simulation are compared in this work for a model of the VNS\cite{BACHMANN2025114796} using both Serpent 2.2.2 \cite{LEPPANEN2015142} and OpenMC 0.15.2 \cite{ROMANO201590}. The production of a variety of medical isotopes in a dedicated irradiation facility in the inboard blanket, as well as the production of \ce{^{60}Co} using \ce{^{59}Co} plates in the outboard blanket are simulated using Serpent 2.2.2.

\section{Codes}
Two continuous-energy Monte Carlo transport codes were chosen for comparison in this study. In general, similar results should be expected since most of the underlying principles of both codes are very similar, as has been shown in recent validation work such as \cite{10682518,VALENTINE2022113197, valentine2021benchmarking}. However, the codes differ in the estimators used for detector responses, depending on the thickness and macroscopic cross section of regions \cite{LEPPANEN2017161}.
\subsection{Serpent 2}
Serpent is a continuous energy Monte Carlo neutron and photon transport code. Models are implemented using universe based Constructive Solid Geometry (CSG). Serpent is developed and maintained by VTT in Finland \cite{LEPPANEN2015142}, is well established in the reactor physics community, and distributed via license acquisition. Serpent is OpenMP and MPI parallelizable.
\subsection{OpenMC}
OpenMC, similar to Serpent is a continuous energy Monte Carlo neutron and photon transport code, with models implemented via CSG \cite{ROMANO201590}. OpenMC has the advantage of being a full open source and community-driven code, meaning that the program's capabilities are expanded directly by the\\ community in addition to the developers, whereas Serpent is not open source. OpenMC is MPI and OpenMP parallelizable and was designed for HPC scalability since its inception.%
\section{Model}
\begin{figure}[ht]
\vspace{-6pt}
    \centering
    \begin{subfigure}[t]{0.4\textwidth}
        \centering
        \includegraphics[width=\textwidth]{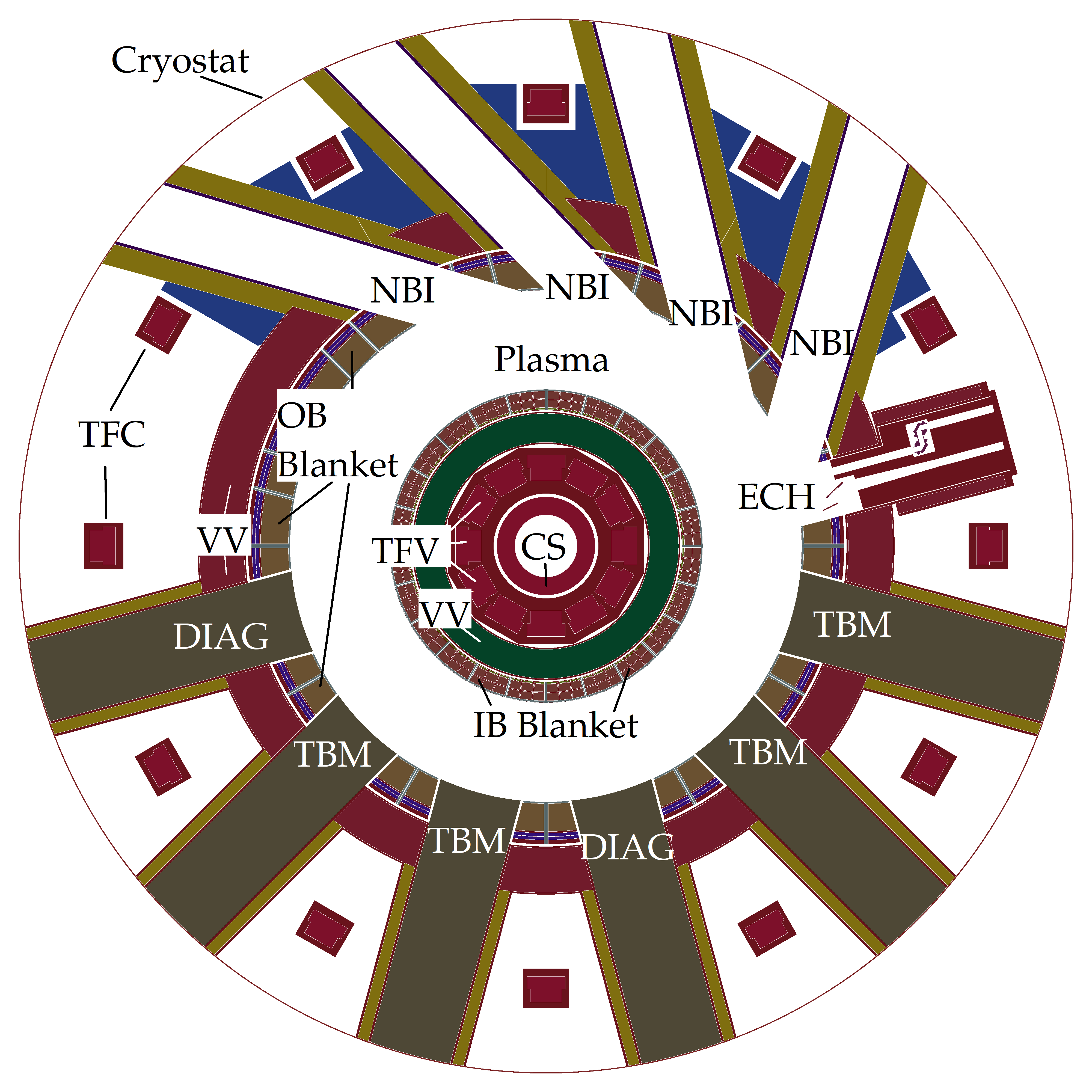}
        \caption{VNS Equatorial Geometry Cross Section}
        \label{fig:xy_geometry}
    \end{subfigure}
    \hfill
    \begin{subfigure}[t]{0.4\textwidth}
        \centering
        \includegraphics[width=\linewidth]{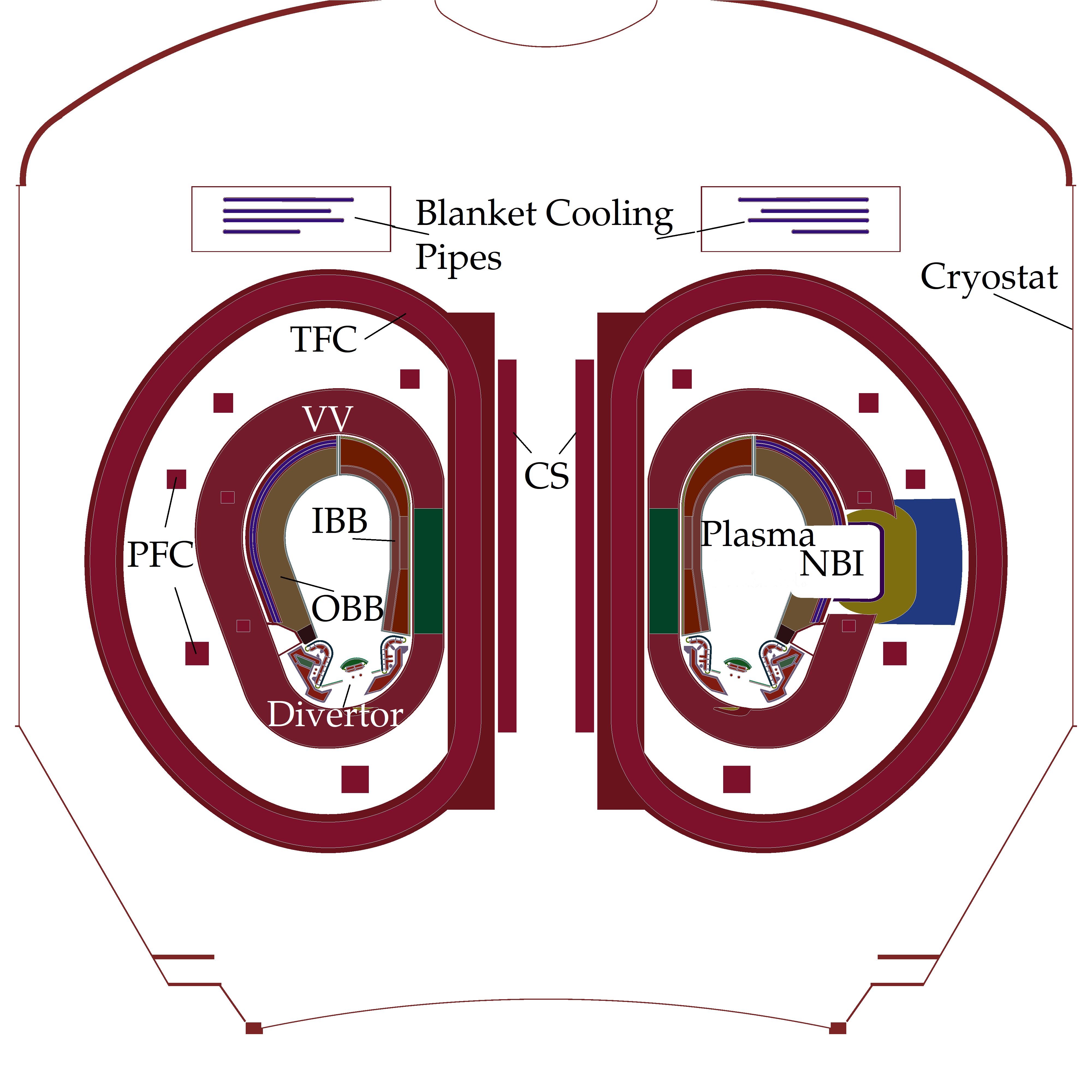}
        \caption{VNS Side Profile Geometry Cross Section}        \label{fig:xz_geometry}
    \end{subfigure}
    \label{fig: geometry}
    \caption{VNS Overview: Neutral Beam Injector (NBI), Inboard Blanket (IBB), Outboard Blanket (OBB), Test Blanket Module (TBM), Diagnostic Modules (DIAG) Plasma,, Vacuum Vessel (VV), Central Solenoid (CS), Toroidal and Poloidal Field Coils (TFC's and PFC's), Blanket Cooling Pipes, Divertor, and Cryostat}
\vspace{-6pt}
\end{figure}
The models were translated independently from an MCNP6 CSG model produced by EUROfusion \cite{Leichtle20062025}. A python script was developed, translating the MCNP model geometry, materials, and the plasma neutron source of the VNS from MCNP to Serpent. The resultant Serpent model was thoroughly checked for geometry errors such as cells and cell overlap using the built-in geometry plotter. An annotated reactor geometry details is to be found in \Cref{fig: geometry}. The only discrepancy found in the model from the MCNP case, was the source energy distribution. In MCNP the source neutron energy follows a Gaussian distribution, which would require modifying the source code of Serpent to implement. Instead a 20-bin weighted energy mesh following the Gaussian was used to accurately sample the energy distribution of source neutrons. The OpenMC model was also scripted from the MCNP model, but using the ``openmc\textunderscore mcnp\textunderscore adapter'' tool developed by \cite{openmc_mcnp_adapter}, for the geometry and materials. A custom script was made to translate the source mesh from MCNP to OpenMC as source translation has not been implemented in the adapter yet. The final model was checked for correct material definitions and geometry errors until the OpenMC model of the VNS was confidently ensured.
\section{Model Comparison Results}\label{sec:Model_Comp}

\begin{figure}[htp]
    \centering
        \begin{subfigure}[t]{=0.38\textwidth}
        \centering
        \includegraphics[width=0.9\textwidth]{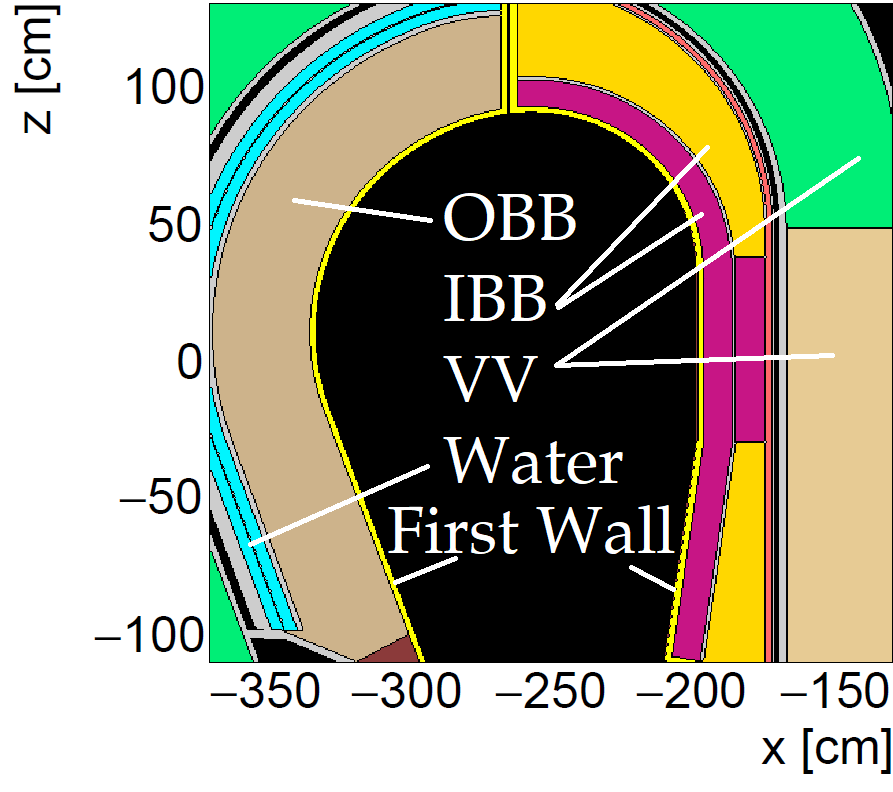}
        \caption{Geometry: In and out board blankets (IBB, OBB), vacuum vessel (VV), water, and first wall}
        \label{fig:xz_blanket_geometry}
    \end{subfigure}\quad  
    \begin{subfigure}[t]{=0.38\textwidth}
        \centering
        \includegraphics[width=0.9\textwidth]{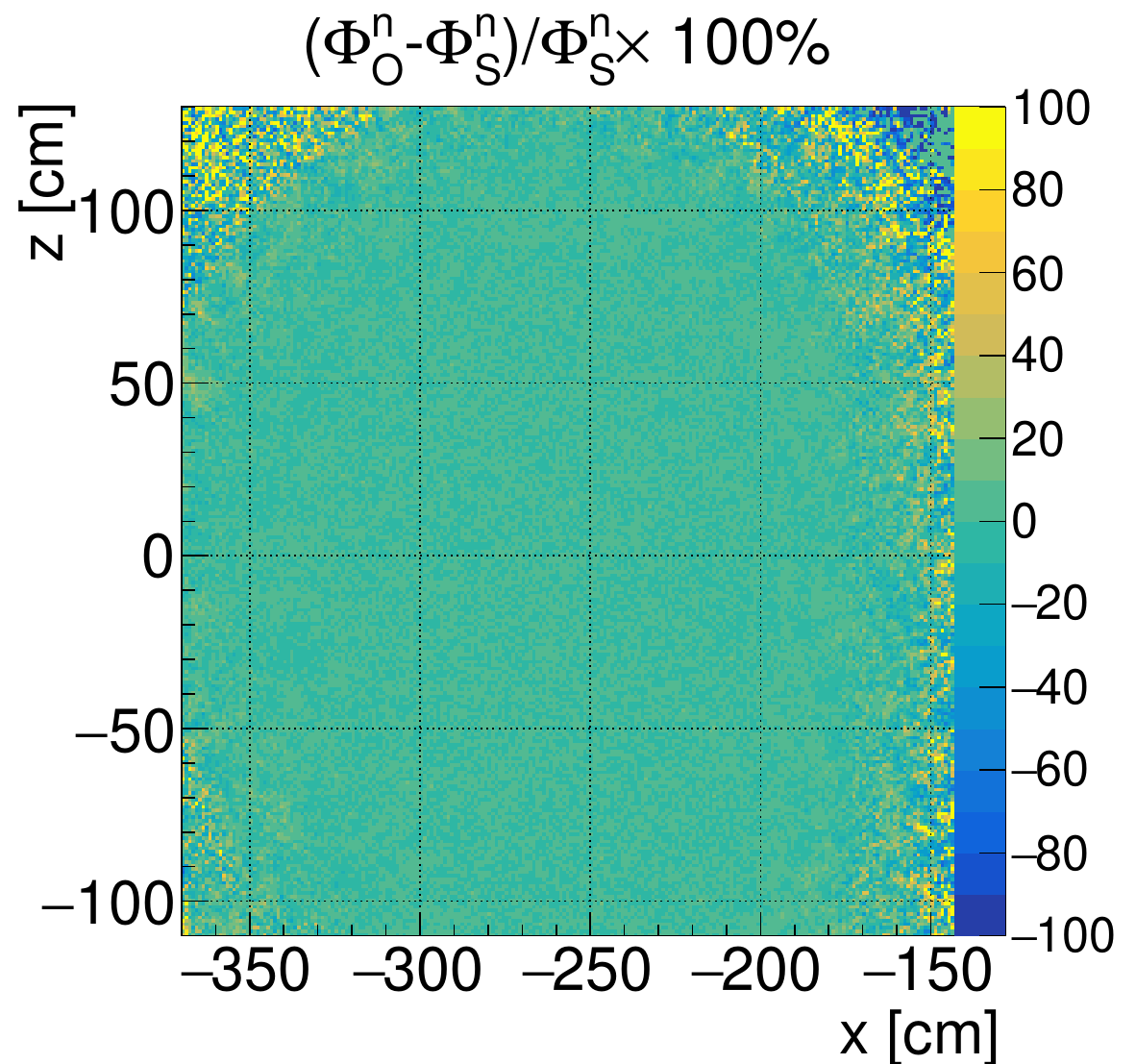}
        \caption{OpenMC-Serpent with surface tracking neutron flux relative difference}
        \label{fig:xz_STnflux_ratio}
    \end{subfigure}\\
    \begin{subfigure}[t]{=0.38\textwidth}
        \centering
        \includegraphics[width=0.9\textwidth]{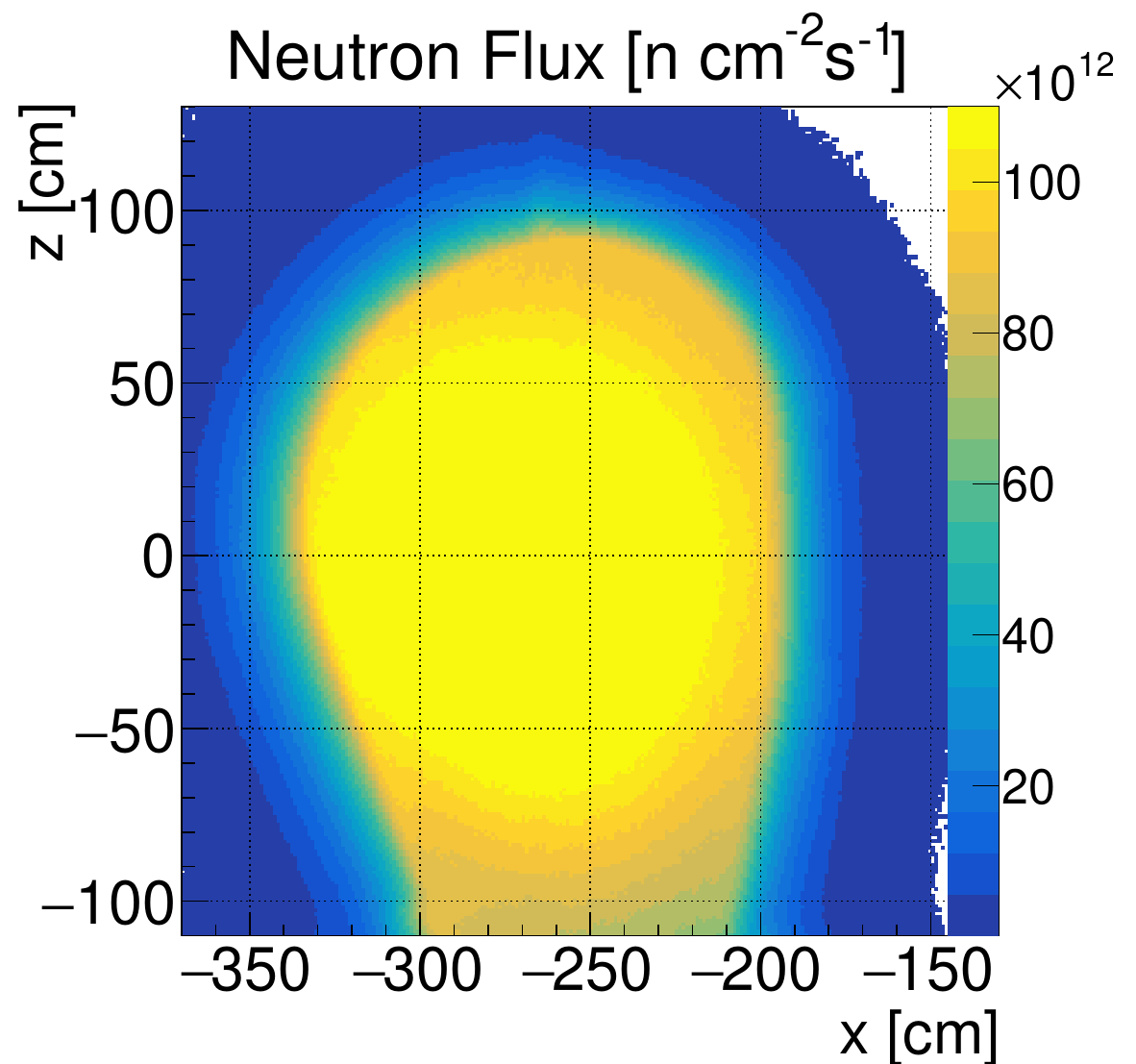}
        \caption{OpenMC neutron flux}
        \label{fig:xz_nflux}
    \end{subfigure}\quad
    \begin{subfigure}[t]{=0.38\textwidth}
        \centering
        \includegraphics[width=0.9\textwidth]{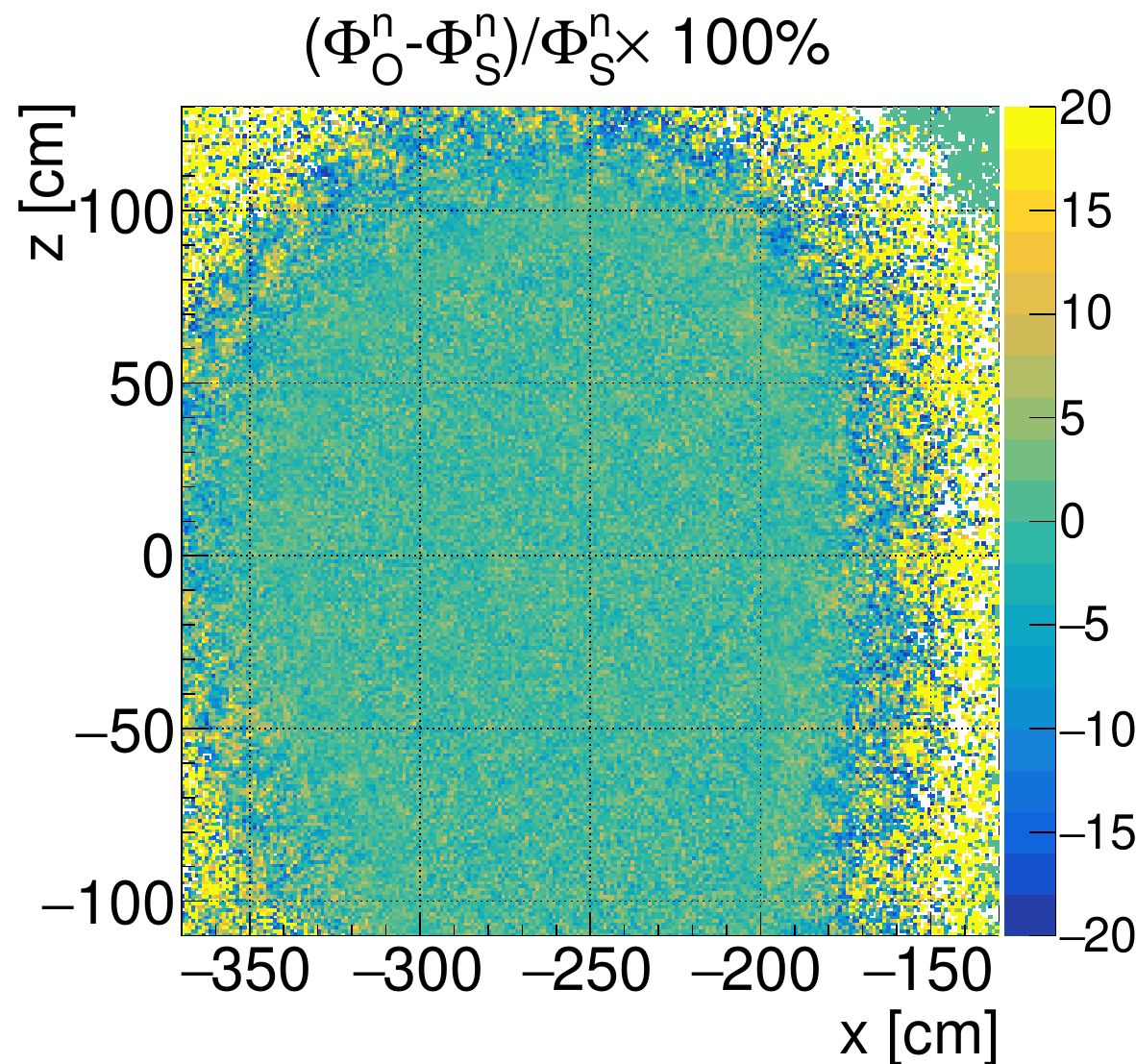}
        \caption{OpenMC-Serpent neutron flux relative difference}
        \label{fig:xz_nflux_ratio}
    \end{subfigure}\\
    \begin{subfigure}[t]{=0.38\textwidth}
        \centering
        \includegraphics[width=0.9\textwidth]{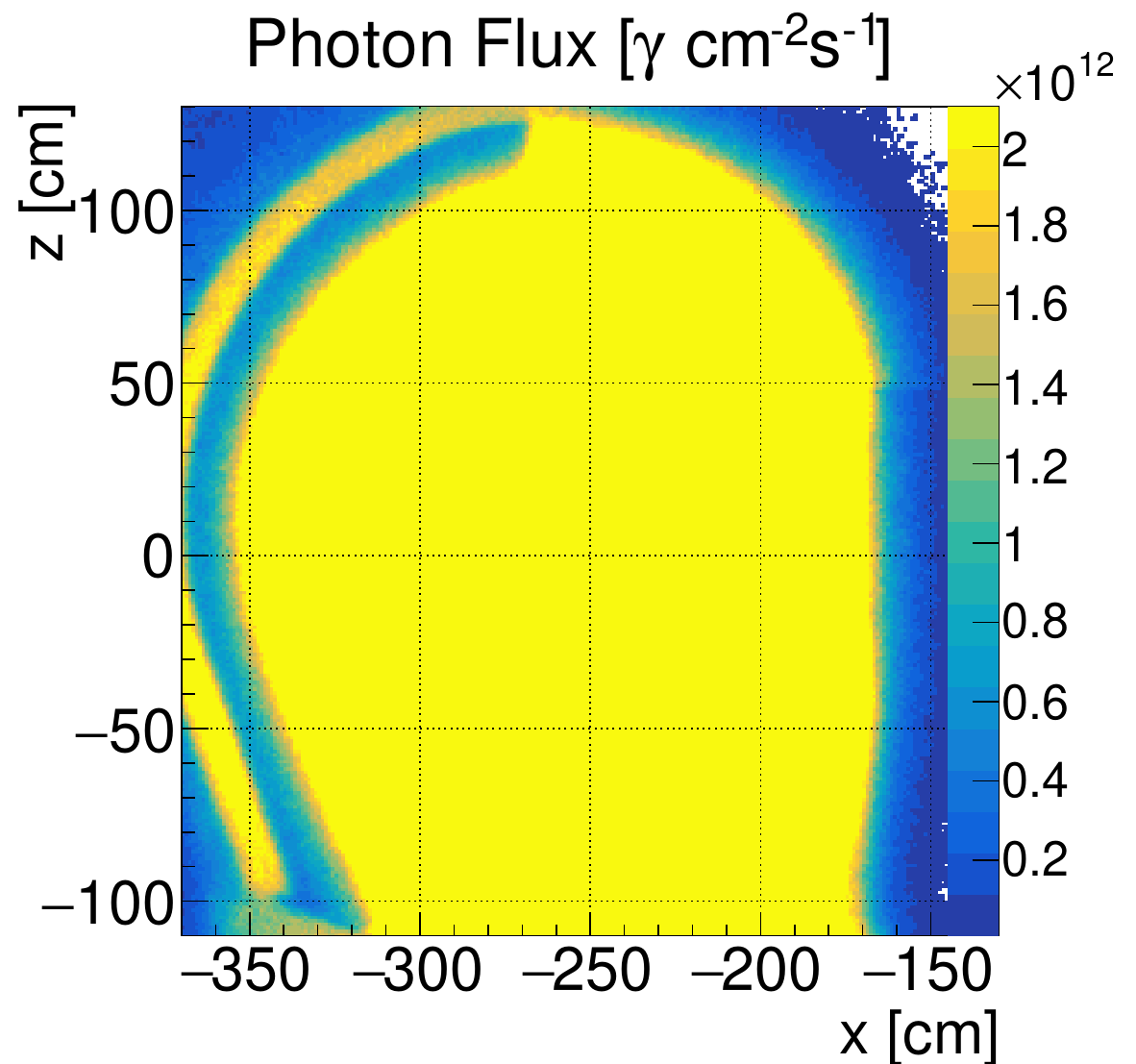}
        \caption{OpenMC photon flux}
        \label{fig:xz_pflux}
    \end{subfigure}\quad
    \begin{subfigure}[t]{=0.38\textwidth}
        \centering
        \includegraphics[width=0.9\textwidth]{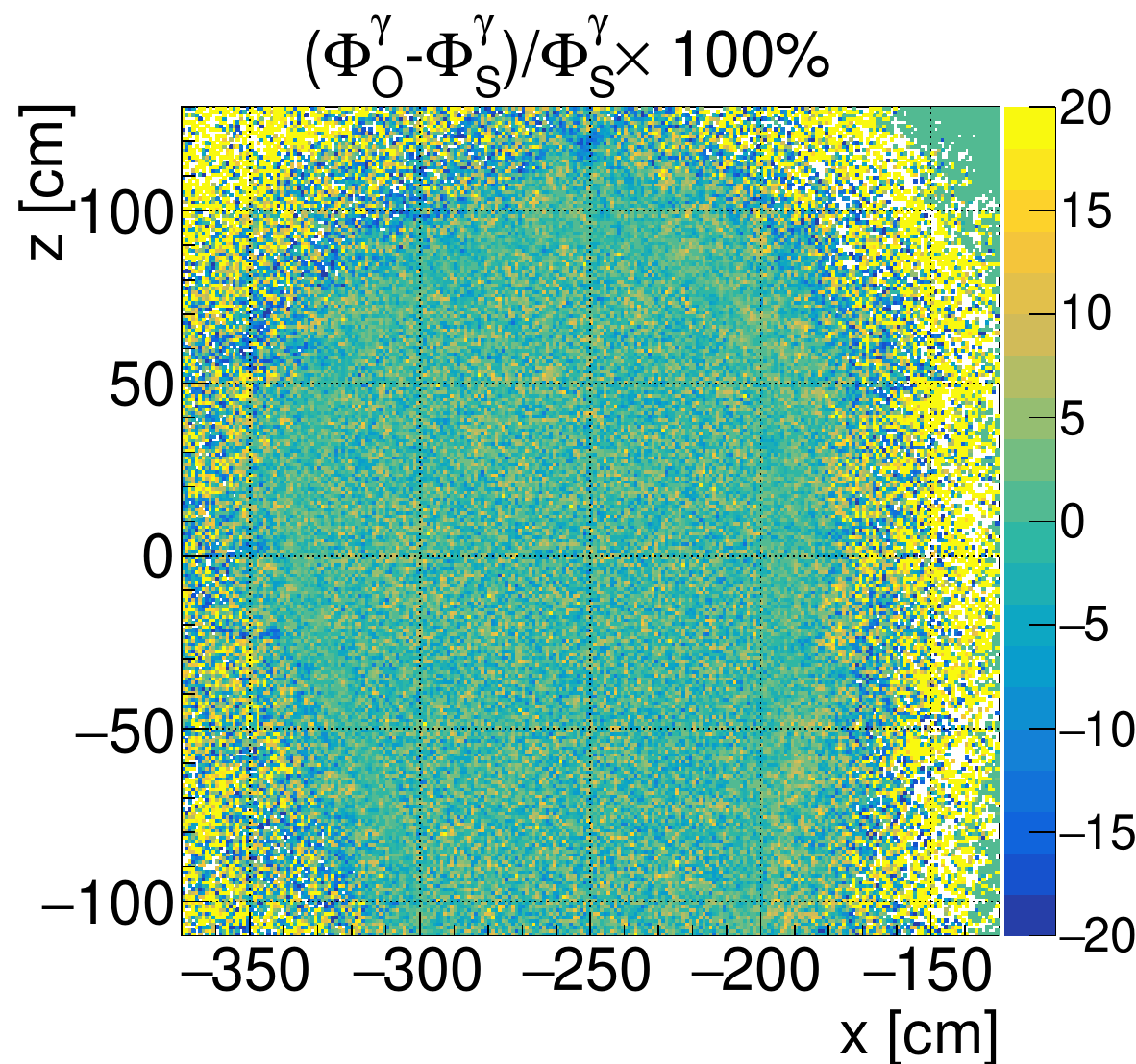}
        \caption{OpenMC-Serpent photon flux relative difference}
        \label{fig:xz_pflux_ratio}
    \end{subfigure}
    \label{fig:xz_flux}
    \caption{OpenMC and Serpent (``O'' and ``S'' subscripts) neutron and photon flux in cut of vacuum vessel}
\end{figure}

\begin{figure}[ht]
    \centering
	\begin{subfigure}[t]{0.38\textwidth}
        \centering
        \includegraphics[width=0.9\textwidth]{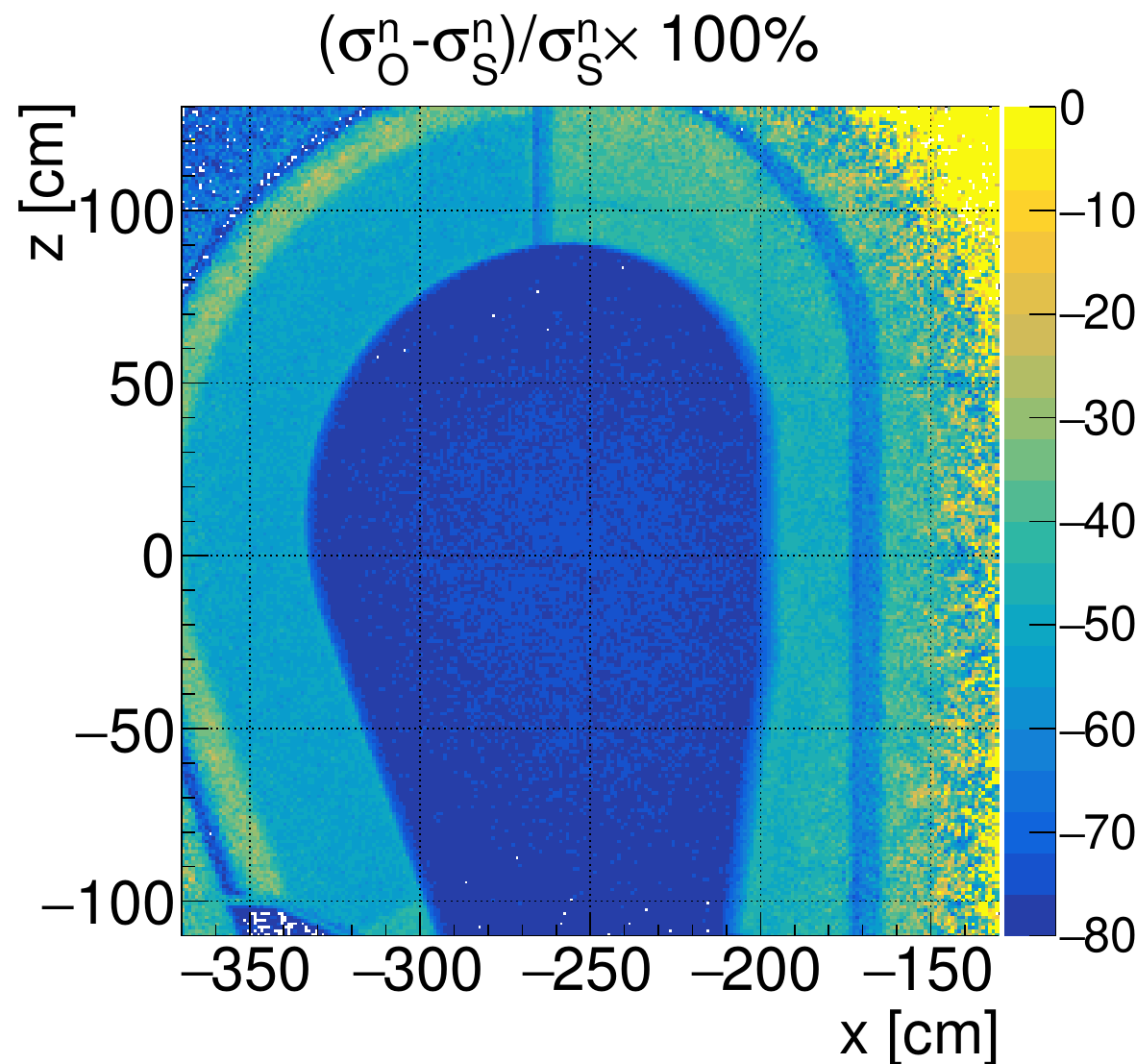}
        \caption{OpenMC-Serpent neutron flux standard error relative difference}
        \label{fig:xz_nflux_error}
    \end{subfigure}\quad
	\begin{subfigure}[t]{0.38\textwidth}
        \centering
        \includegraphics[width=0.9\textwidth]{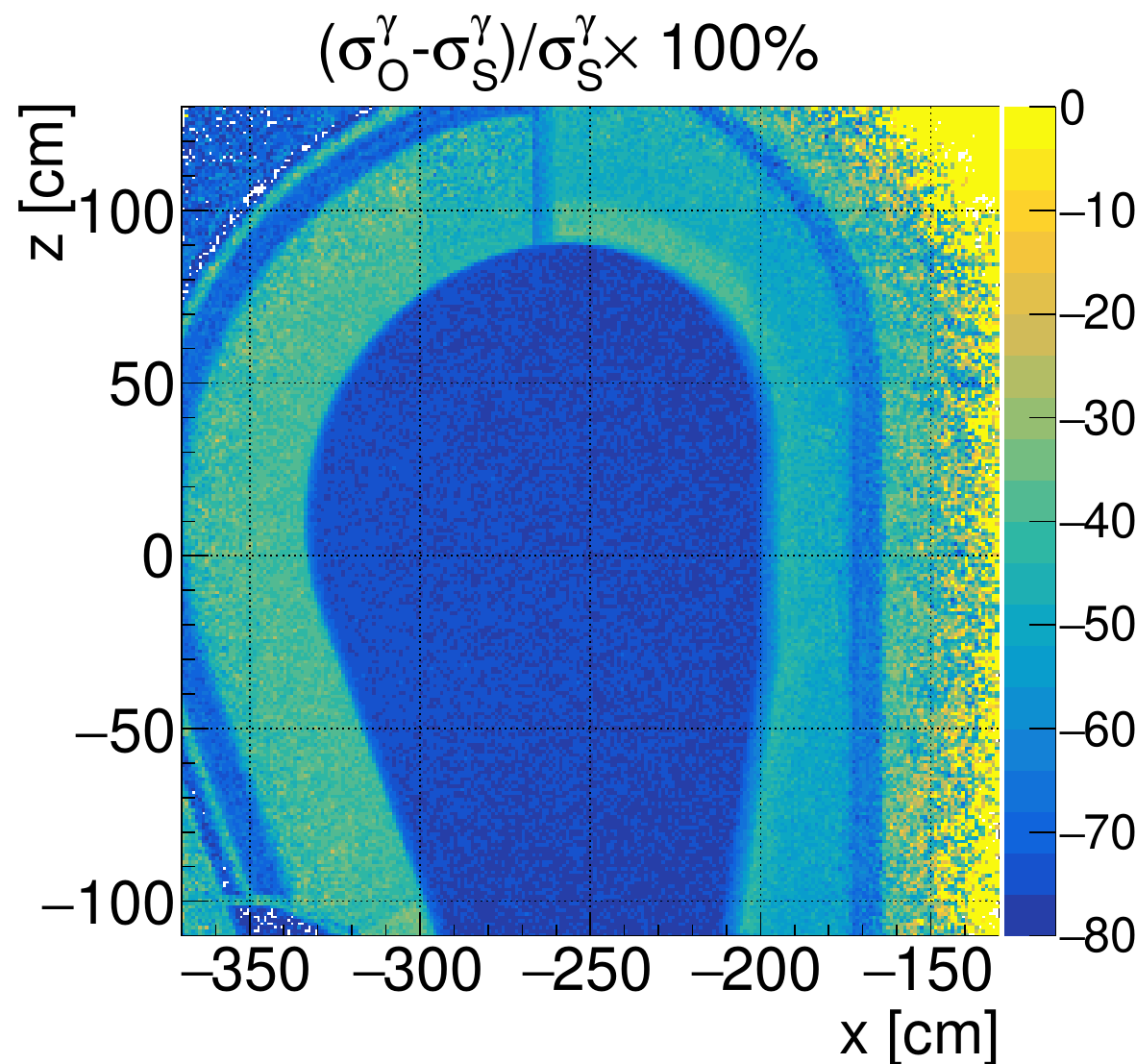}
        \caption{OpenMC-Serpent photon flux standard error relative difference}
        \label{fig:xz_pflux_error}
    \end{subfigure}
    \label{fig:xz_flux_error}
    \caption{OpenMC and Serpent (``O'' and ``S'' subscripts) neutron and photon flux relative error in cut of vacuum vessel}
\end{figure}
The differences in VNS related calculations, were performed for a mapping of the neutron and gamma fluxes, ($n,T$) and ($n,2n$) reaction rates to verify in detail how sensitive the output results of Serpent and OpenMC are. Specific regions are inspected within the VNS to estimate possible discrepancies between codes for reaction rates, particle flux values, and the quality of statistical uncertainty achieved in the calculations. The codes were each run with \SI{1e8}{neutron} histories divided into 1,000 batches in external source mode using the well established and readily available ENDF/B-VII.1  \cite{Chadwick20112887} evaluated nuclear data library as a common nuclear data basis for code comparison. Neutron and photon flux detector responses were compared for a segment of the VNS vacuum vessel. A geometry plot of the tallied region is shown in \cref{fig:xz_blanket_geometry}, including a cooling pipe, IB and OB blankets, the first wall, and \ce{TiH}$_2+$\ce{W} shielding, of the reactor for OpenMC and Serpent. The models resulted in very similar neutron and photon fluxes, agreeing to $0.086\%$ and $0.028\%$ within statistical uncertainties for the average values of \cref{fig:xz_nflux_ratio} and \cref{fig:xz_pflux_ratio}, no strong difference predicted between OpenMC and Serpent. This is even better agreement than the same calculation, but with Serpent 2.2.0 instead of Serpent 2.2.2  yielded in \cite{c_ehrich_2026_20804012} to be 0.14\% and 1.8\% for neutron and photon fluxes respectively. The central region of the flux difference maps \cref{fig:xz_nflux_ratio} and \cref{fig:xz_pflux_ratio} are within precise agreement, while it can be seen that the peripheral region (further along the poloidial radius) becomes statistically noisier due to fewer neutron tallies. The standard deviation values themselves were also compared (see \cref{fig:xz_nflux_error} and \ref{fig:xz_pflux_error}), and it was found that the uncertainty was lower for OpenMC than Serpent, with the mean flux standard deviation relative difference being 65.5\% and 64.5\% for neutrons and photons respectively. This difference is caused by the large void region, which suffers from fewer tallies in Serpent and has to do with the detector response estimator used. The lower tallying for the photon flux results from the much lower photon population compared to neutrons (see \cref{fig:xz_nflux,fig:xz_pflux}). Upon further inspection, the tally sampling is different between codes (see \cref{fig:xz_nflux_error}) despite cross-code agreement between neutron flux results in \cref{fig:xz_nflux_ratio}. This may be due to the different tally estimators between codes. OpenMC uses a track length estimator (TLE) to sample tallies, where as Serpent uses the collision flux estimator (CFE) for particles without interaction every \SI{20}{cm}, mixed with a track length estimator. The collision flux estimator depends on the collision frequency (and hence macroscopic cross section) of the material. \cref{fig:xz_nflux_error} and \ref{fig:xz_pflux_error} clearly shows that certain regions sample more tallies than others. The void region with no interaction probability, where source neutrons are generated, has about 75\% lower uncertainty in OpenMC's TLE tally collector compared to Serpent. The top left corner is another void region, where a higher statistic can be seen in OpenMC than Serpent, and just before that corner, is a region with much higher material density, and a more similar agreement in sampling can be seen between the codes. Good agreement is observed for the photon flux direct response.
\FloatBarrier

\subsection{Reaction Rate Comparison}
\begin{figure}[ht]
    \centering
    \includegraphics[width=1.0\linewidth]{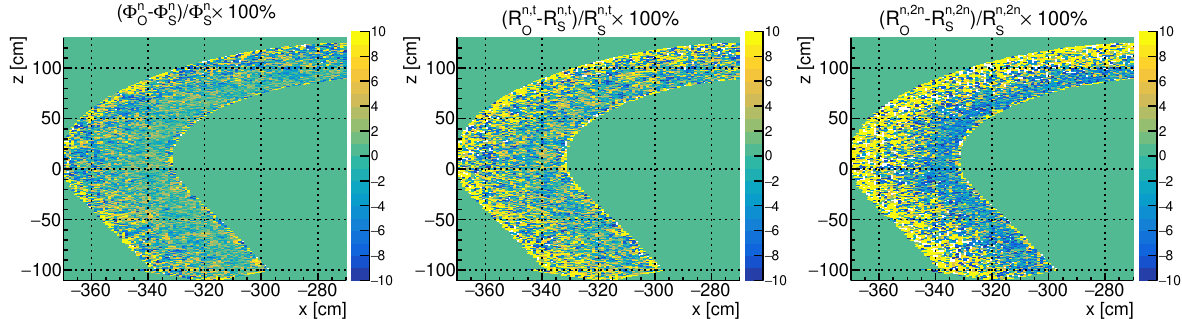}
    \caption{OB blanket OpenMC to Serpent relative difference of neutron flux, ($n,T$), and ($n,2n$) reactions }
    \label{fig:OB_flux_ratio}
\vspace{-6pt}
\end{figure}
\begin{table}[ht]
  \centering
  \caption{OpenMC and Serpent with Hybrid, $\delta$, and Surface Tracking Estimator Total Outboard Blanket Flux, ($n,T$) Reaction Rate, and ($n,2n$) Reaction Rate. The statistical uncertainty of each tally is $\pm0.02\%$}
  \label{tab:integral_reaction_rates}
  \begin{tabular}{lcccc}
    \toprule
    Code & $\Phi^n$ [n$\cdot $cm$^{-2}$s$^{-1}$] &$\Phi^\gamma$ [$\gamma\cdot$ cm$^{-2}$s$^{-1}$] & ($n,T$) [s$^{-1}$]& ($n,2n$) [s$^{-1}$]\\
    \toprule
    Serpent  & $2.69\times10^{13}$ & $4.69\times10^{12}  $ & $4.52\times10^{11}$  & $1.27\times10^{11}$ \\
    \midrule
    OpenMC  & $2.65\times10^{13} $& $4.69\times10^{12} $ & $4.46\times10^{11} $& $1.20\times10^{11} $\\
    \midrule
    $\frac{\text{OpenMC-Serpent}}{\text{Serpent}}$ &  $[-1.6 \pm 1.3]\%$ & $[-0.1 \pm 1.4]\%$ & $[-1.5 \pm 1.5]\%$ & $[-5.4\pm 1.4]\%$\\
    \bottomrule
  \end{tabular}
\end{table}
In addition to the triton production, neutron multiplication reactions i.e. ($n,2n$), are important to compensate a neutron flux depletion due to absorption in the blanket and to ensure sufficient Tritium breeding. These two reactions were tracked in the outboard blanket and compared as a ratio (shown in \cref{fig:OB_flux_ratio} with the OB blanket integral results shown in \cref{tab:integral_reaction_rates}) and come to close agreement. Serpent 2.2.2 systematically estimates slightly higher the tally results than OpenMC 0.15.2, with the largest reaction rate discrepancy found being ($n,2n$) at $[-5.4 \pm 1.4]\%$. Photon flux and the triton production rate differences across codes are within one standard deviation uncertainty, and neutron flux difference is only 0.3\% larger than uncertainty. 

Two different tally estimators are used by default, Serpent uses a collision flux estimator (CFE), which samples virtual collisions every \SI{20}{cm} in addition to physical collisions, and is most effective in low mean-free-path regions \cite{LEPPANEN2017161}. OpenMC uses the more versatile track length estimator (TLE), which records the length traveled by a particle between surface crossings and uses the region's macroscopic cross section to estimate reaction rates. As the geometry simulated is a mix of vacuum, and medium to strong absorbers, the low macrosopic cross sections regions may bias the simulation in CFE cases. Increasing the collision frequency by a factor of 10 was tested to see if more virtual collisions could yield closer results, but it did not. As OpenMC uses surface tracking (ST) \cite{ROMANO201590}, the Serpent simulation was rerun using surface (see \cref{fig:xz_STnflux_ratio}) and delta tracking ($\delta$-T) separately instead of the default hybrid delta-surface tracking, which uses ST in regions with collision efficiency below $90\%$ real collisions. For $\delta$-T, the collision efficiency was very low ($1.8\%$ for photon and $3.9\%$ for neutron), while ST yielded higher values ($88\%$ for photon and $71\%$ for neutron). The Serpent surface tracking flux results agreed with OpenMC in the peripheral region with more than in hybrid tracking mode (seen in \cref{fig:xz_STnflux_ratio}) likely due to higher sampling in the peripheral region, while $\delta$-T yielded results similar to hybrid tracking. Employment of $\delta$-T cost time, leading to $48\%$ longer computation than  than hybrid mode, likely due to low sampling efficiency, while ST only took $12\%$ longer than hybrid mode. The best practice case for VNS geometry therefore is to use surface tracking in Serpent.
\subsection{Spectral Flux Comparison}
\begin{figure}[ht]
    \centering
    \begin{subfigure}[t]{0.35\textwidth}
        \centering
        \includegraphics[width=\textwidth]{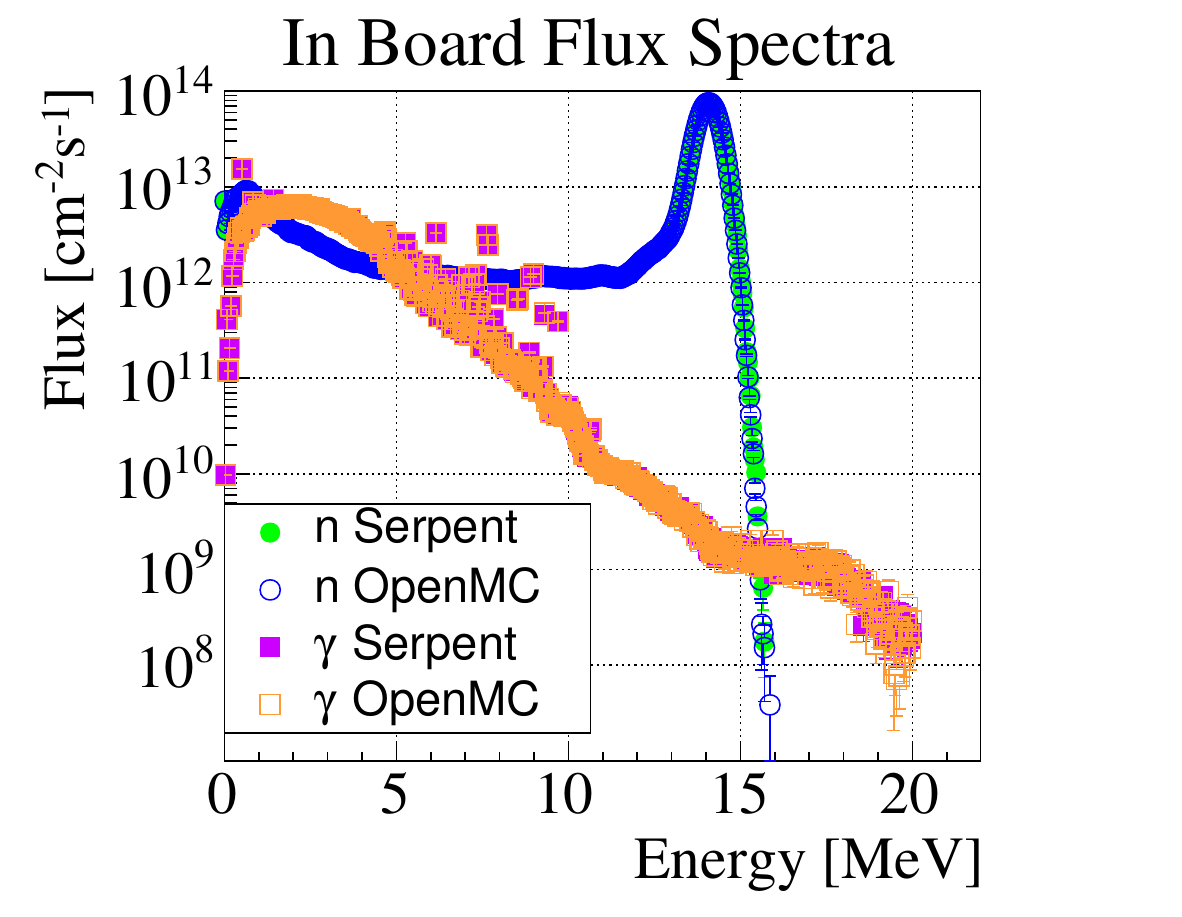}
        \caption{IB blanket Neutron and Photon flux spectra}
        \label{fig:IB_flux_spectra}
    \end{subfigure}\quad
    \begin{subfigure}[t]{0.35\textwidth}
        \centering
        \includegraphics[width=\textwidth]{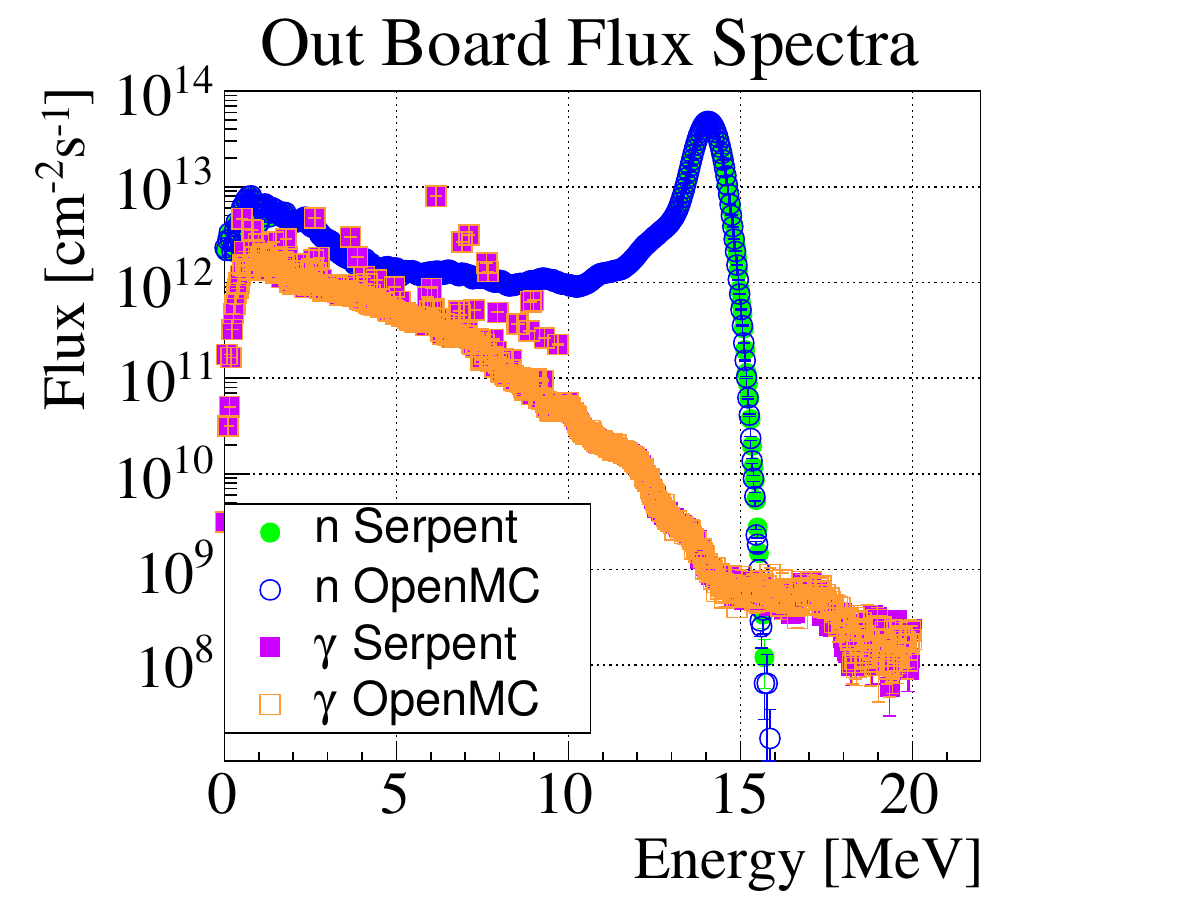}
        \caption{OB blanket Neutron and Photon flux spectra}
        \label{fig:OB_flux_spectra}
    \end{subfigure}\\
    \begin{subfigure}[t]{0.35\textwidth}
        \centering
        \includegraphics[width=\textwidth]{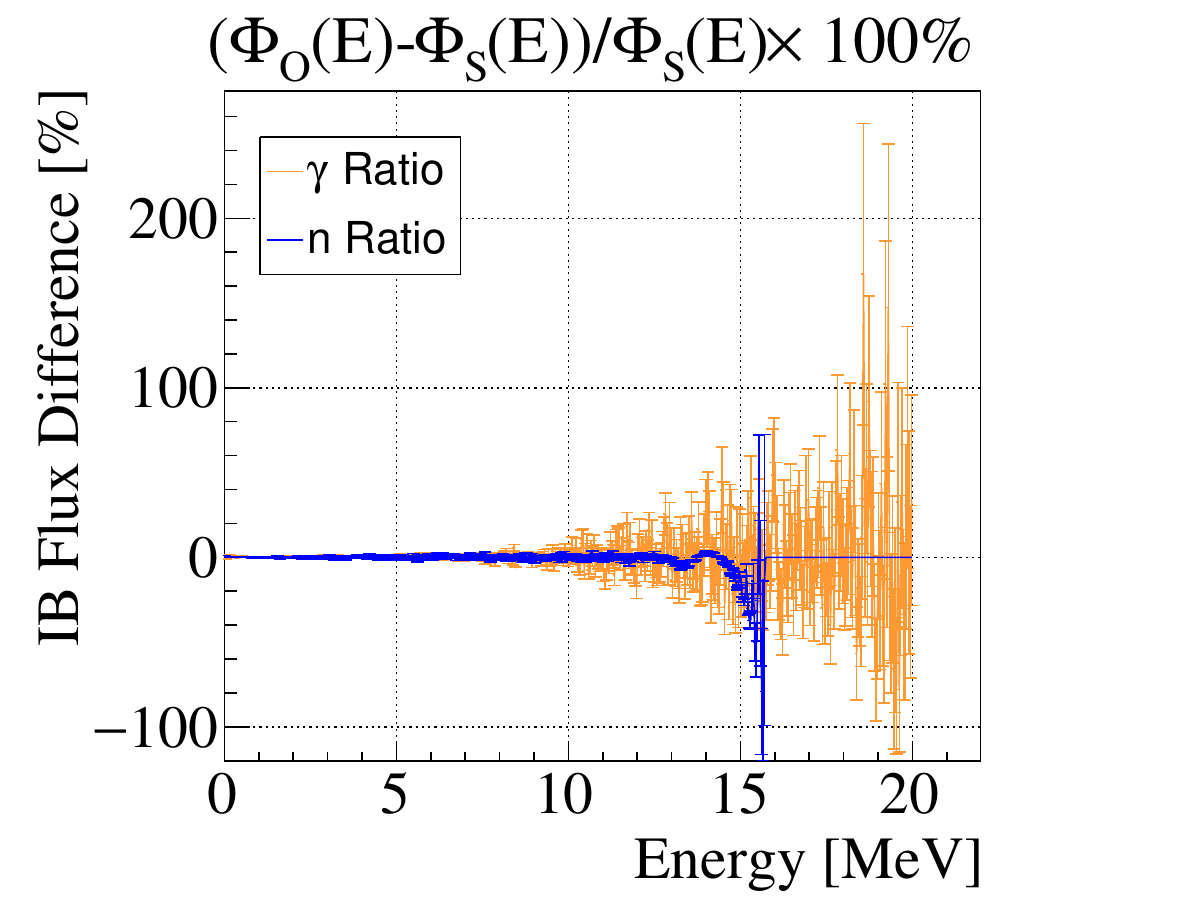}
        \caption{IB blanket Neutron and Photon flux spectra ratios}
        \label{fig:IB_flux_spectra_ratio}
    \end{subfigure}\quad
    \begin{subfigure}[t]{0.35\textwidth}
        \centering
        \includegraphics[width=\textwidth]{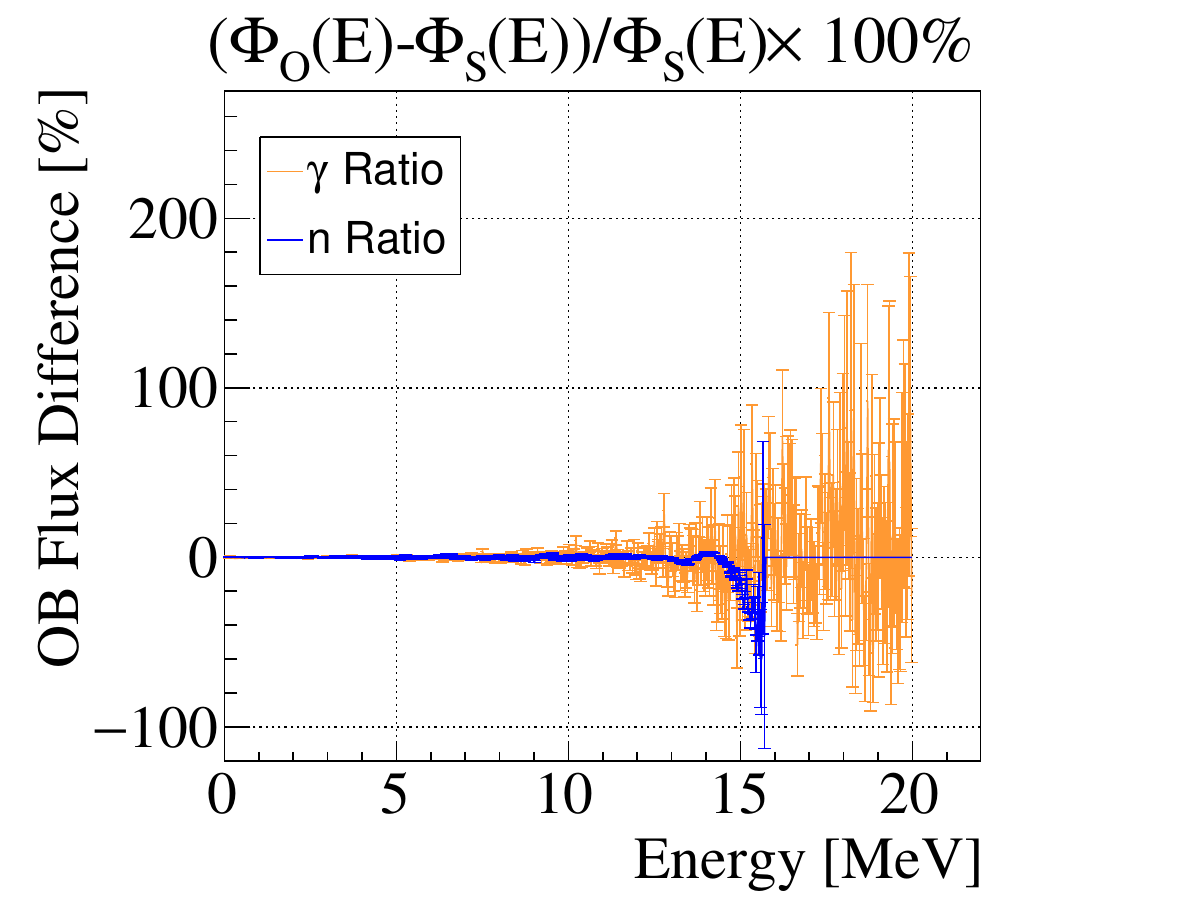}
        \caption{OB blanket Neutron and Photon flux spectra ratios}
        \label{fig:OB_flux_spectra_ratio}
    \end{subfigure}
    \label{fig:flux_spectra}
    \caption{Flux spectra comparison in blanket regions}
\end{figure}

The neutron and photon flux of IB and OB blanket regions were plotted for OpenMC and Serpent in \cref{fig:IB_flux_spectra,fig:OB_flux_spectra}. Each detector uses 500 evenly spaced energy bins, from 0 to \SI{20}{MeV} for neutrons, and from \SI{7.5}{keV} to \SI{20}{MeV} for photons. On both IB and OB blankets, except for in the high energy photon, and to a lesser extent neutron regimes, which suffer from low sampling, overall good agreement can be seen. One can see the \SI{14}{MeV} Gaussian overlaps nearly perfectly, justifying the source energy distribution used. In the high energy ($>$\SI{14}{MeV}), low sampling and flux tail of the distribution, Serpent calculates higher values than OpenMC, but are comparable with the high error that increases with energy. On the higher sampling ($\leq$\SI{14}{MeV}) side of the distribution agreement is seen to at most 3.2\% difference between OpenMC and Serpent for both IB and OB blankets. Individual resonances $<$\SI{10}{MeV} in the spectra visibly match. It can be seen from \cref{fig:OB_flux_spectra_ratio} that the OpenMC IB and OB photon flux spectra agree within statistical uncertainties, an agreement that improved from Serpent 2.2.0 results in \cite{c_ehrich_2026_20804012}.  
\subsection{Simulation Time Comparison}
\begin{figure}[ht]
    \centering\includegraphics[width=0.5\textwidth]{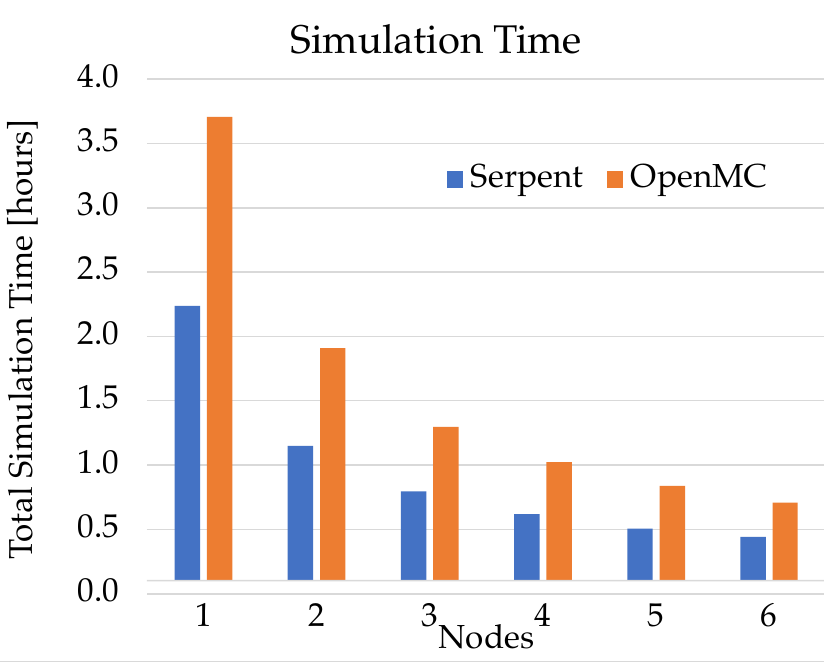}           
    \captionof{figure}{Simulation times on HPC (note different compilers were used for Serpent and OpenMC)}
    \label{fig:sim_time_upscale}
\end{figure}
A comparison between total simulation times of the Serpent and OpenMC, in photon-neutron coupled simulation mode on the Leibniz Rechenzentrum HPC was performed by scaling the number of nodes used. Each node is composed of a 160 Intel Xeon Platinum 8380 Ice Lake CPU threads. In each case, Serpent (using default particle tracking) was between 1.6 and $1.7\times$ faster than OpenMC, as plotted in \cref{fig:sim_time_upscale}. This result is likely partially due to the speed advantage of the CFE which samples fewer tallies than the TLE employed by OpenMC. This result is consistent with \cite{VALENTINE2022113197} which also presents Serpent results as faster than OpenMC for neutron-photon coupled transport. Further comparison to \cite{VALENTINE2022113197} was made using neutron only mode, where \cite{VALENTINE2022113197} found OpenMC  $1.85\times$ faster than Serpent. In VNS geometry, neutron only performance was $1.6\times$ faster in OpenMC than Serpent on one node. In addition to differences between codes, this installation of Serpent may have a systematic speed advantage over OpenMC up to $\approx 1.3\times$ \cite{10.1007/978-3-031-95130-5_14} due to it using mpiicx whereas OpenMC used a GCC compiler.
\FloatBarrier
\section{Medical Isotope Generation} 
The aim of this work is to estimate medical isotope yields in the VNS by simulated isotope production. The VNS Serpent model was modified to arrange a dedicated space behind the first wall of the IB blanket, wrapping $2\pi$ about the toroidal axis. This space is reserved to accommodate thin tubes of \SI{2}{cm} diameter filled with capsules containing precursor materials. 

\begin{figure}[htb]
    \centering
    \begin{subfigure}[t]{=0.45\textwidth}
    	\centering
        \includegraphics[width=\linewidth]{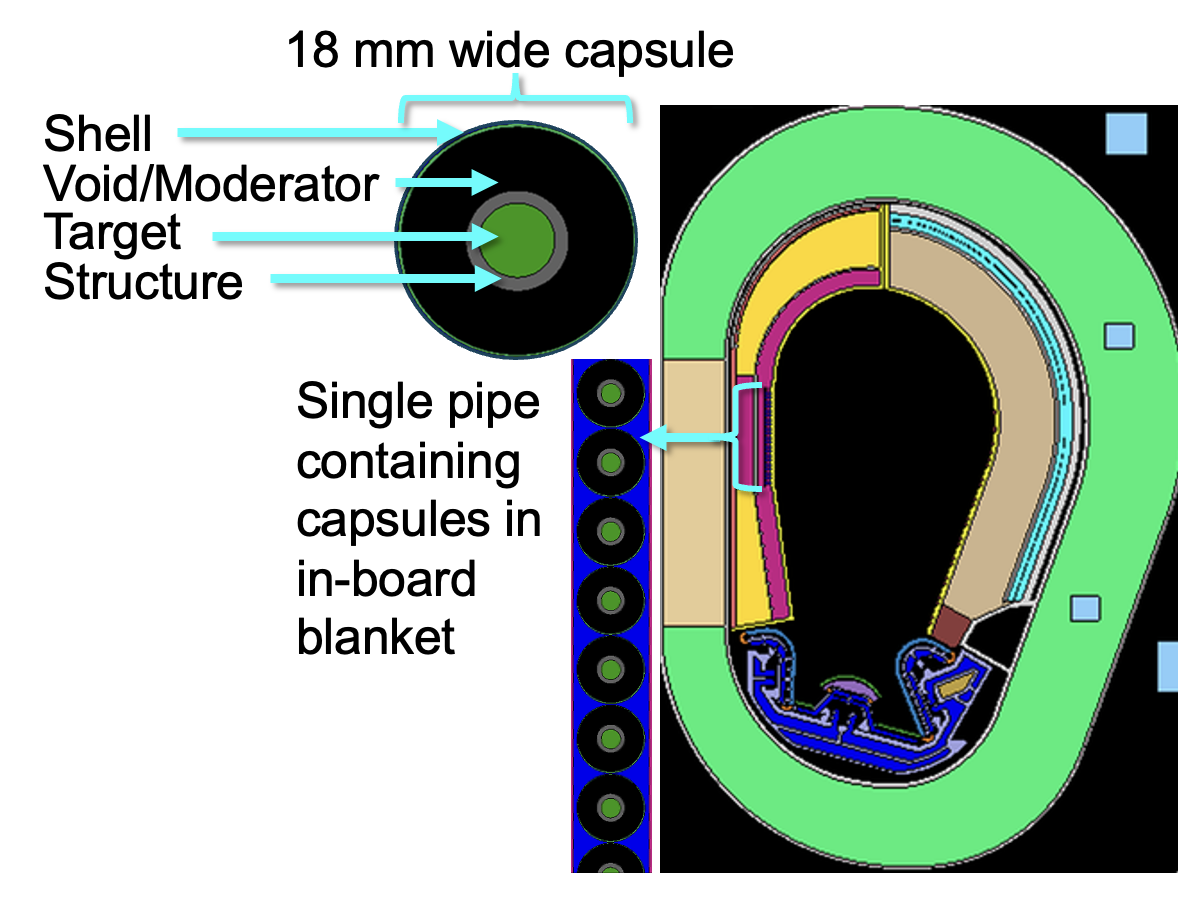}
        \caption{Serpent model of VNS irradiation facility with example capsules}
        \label{fig:caps-model}
    \end{subfigure}\quad
    \begin{subfigure}[t]{=0.45\textwidth}
    	\centering
        \includegraphics[width=\linewidth]{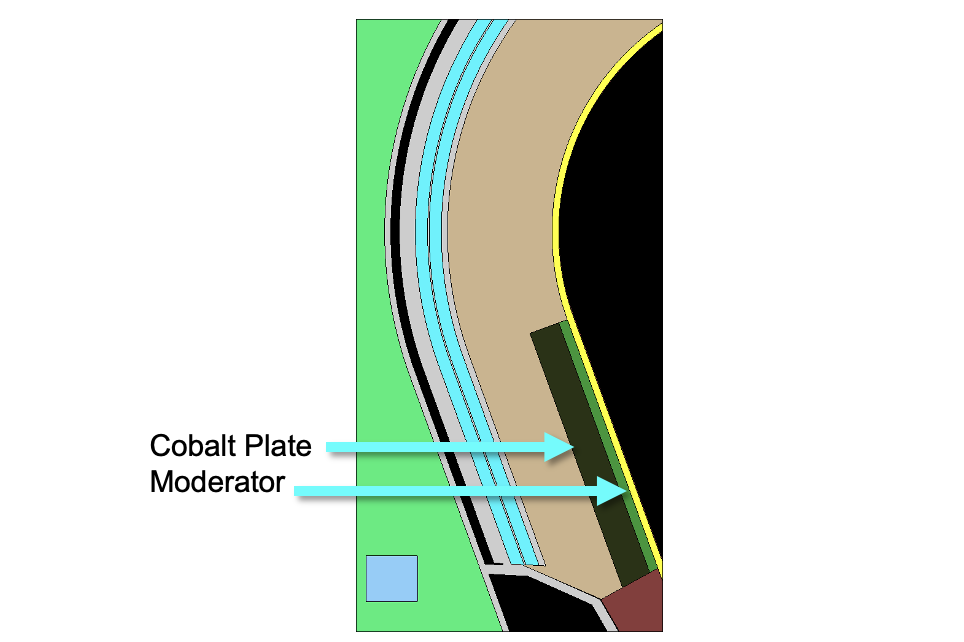}
        \caption{Example Serpent model of VNS Cobalt plate behind moderator material}
        \label{fig:cobalt-model}
    \end{subfigure}
    \caption{Serpent model of VNS irradiation facilities}
    \label{fig:irrad-models}
\end{figure}

The idea of medical isotope production in fusion is not a new concept \cite{Engholm01111986} and is in fact of continued interest in the fusion community \cite{parisi2025productionhighspecificactivityradioisotopesusing,cho2024making}. The selection of isotope production reaction channels was supported using a tool developed alongside this project in collaboration with the Bundesargentur für Sprunginnovation (SPRIND) \cite{sprind_ein1091}. The tool can filter isotope production reactions by particle source facilities, such that reactions suitable for \SI{14.1}{MeV} fusion neutrons could be identified and exploited.

\subsection{Modeled Irradiation Facilities}

The study is based on a reference design of the target deployment and recovery system developed and kindly provided by KFT Konzepte, called ``RIGS'' \cite{kft_radioisotopes}. The system was developed specifically to expose target materials to the neutron flux generated in the plasma of a fusion based Volumetric Neutron Source to cause the generation of radioactive isotopes. The RIGS system makes use of the neutrons that are otherwise foreseen to be absorbed on the inboard side, and it deploys and recovers precursor materials in the inboard blankets without compromising their neutron absorbing function.

The radioisotope recovery system is a closed loop of circulating water, that consists of auxiliary components operating the loop that are located above the bioshield roof and of a pipeline that is routed in several bends behind the plasma-facing first wall of a single blanket. 24 identical systems are considered in this paper to be operated in parallel connected to the 24 inboard blankets. The pipe connects the blanket with the auxiliary system and penetrates the primary vacuum vessel. Inside the pipe, spherical capsules are submerged in the liquid water and are transported through the forced water flow. The inner diameter of the pipe is assumed here to be \SI{20}{\milli\meter}, that of a spherical capsule to be \SI{18}{\milli\meter}. The pipe length of a single system is \SI{81.5}{\meter}, \SI{50}{\meter} of which located behind the blanket first wall where the capsules are exposed to the neutron flux. Capsules in the remaining part of the loop are assumed not to be exposed to neutrons. Each capsule is assumed to carry \SI{1.1}{\gram} of target material. Consequently, in each system 2,763 capsules are exposed to neutron flux at any point in time i.e., 2,763 $\times$ \SI{1.1}{g} = \SI{3.04}{\kilo\gram} of target material. The total amount of target material exposed to neutron flux in all 24 at a given time  systems is \SI{72.9}{\kilo\gram}, which is the basis of this study.

Since the auxiliary systems of the loop, including the system to refill new spherical capsules and recover capsules that completed the irradiation process, are located in areas protected from excessive neutron and gamma radiation, they are well accessible during plasma operation or during plasma shutdown. The preparation of the spherical capsules with target materials and their recovery after irradiation is carried out in the hot cell building. The transfer of the spherical capsules from the area outside the bioshield and the aforementioned hot cells makes use of the transfer system implemented in the nuclear fusion facility for the remote replacement of the blankets.

The capsules were modeled in Serpent as concentric spheres of target material, followed by Al6061 cladding/structural material, followed by vacuum or moderator, and finally a stainless steel 316 L shell submerged in pipe water as shown in \cref{fig:caps-model}. Al6061 was chosen as a material due to its neutronic transparency, radiation hardness, and favorable mechanical properties \cite{alexander1992irradiation}. Similarly, the delta phase of Zirconium Hydride: \ce{ZrH_{1.6}} was chosen as the primary moderator because of its relative radiation hardness and high slowing down power \cite{snead2022development}, and is compared against \ce{ZrH_{2}} -a phase with higher Hydrogen packing, but less Hydrogen retention stability, and against water as moderators. 

The capsules are assumed to be only exposed to flux 61\% of the time, due to the loop segments outside of the irradiation zone. To reflect this in the model the flux is scaled down to 61\% of the original ca. $1\times10^{14}$ n$\cdot$ cm$^{-2}$s$^{-1}$ IB blanket flux, and one depletion zone was shared across all capsule targets in the system. The flux scaling approximation is justified because the time spent outside the loop is small compared to the irradiation time scale and the neutron fluency and decay time are preserved. The usage of one shared depletion zone is justified because every capsule will experience the flux in all locations in the irradiation facility loop.

A further simplification by the Serpent model is that  21.5\% of the available irradiation volume is modeled, or \SI{15.7}{kg} out of \SI{72.9}{kg} of target material irradiated at a time. The neutron wall loading and total \SI{118}{kg} of target material (exposed to and not exposed to flux) are then used to extrapolate the total predicted yield. The flux is dependent on the poloidial position about the plasma, which the irradiation facility is designed symmetrically around, so the number of reactions is scaled by the radial neutron wall load. This simplification assumes that the spectrum is poloidially independent. A final extrapolation factor of 6.33 is calculated and applied to the raw Serpent isotope yields, which accounts for the volumes not modeled in Serpent. In the simulated cases of this study, the VNS is loaded fully with only capsules containing the same target material, however in principle capsules with different targets could also be loaded into the facility together and irradiated simultaneously. Further capsule design optimization could be performed.

In addition to the in-board capsule irradiation facility, an out-board Cobalt plate was modeled for irradiation shown in  \cref{fig:cobalt-model}.The out-board plate cases were simulated separately from the in-board irradiation facility cases. The general design consists of a truncated conical plate toroidially wrapped around the VNS Tokamak out-board blanket in $2\pi$, modeled primarily below the ports with small gaps made for their accommodation. Further design refinement is foreseen, depending on availability of the individual out-board blanket modules. Four different cases were initially considered, including \SI{1}{cm} and \SI{10}{cm} thick cobalt unmoderated plate cases, and \SI{1}{mm} and \SI{10}{cm} thick cases moderated behind a \SI{3}{cm} thick \ce{ZrH_{1.6}} plate between the Cobalt plate and the plasma.

\subsection{Predicted Medical Isotope Yields}
Burnup simulations using the predictor corrector method with incremental burnup step sizes ranging from 0.1 to 15 days and the total number of steps ranging from 12 to 15 depending on the isotope were performed. The simulated irradiations were carried out using $10^{8}$ neutron histories each, analog neutron-photon coupling, and default hybrid tracking mode. The mature and extensively validated ENDF/B-VII.1 \cite{Chadwick20112887} nuclear data library was chosen and used as a baseline reference unless JEFF 4.0 \cite{jeff40_neutron} is specified. JEFF 4.0 was used for sensitivity analysis to nuclear data, as certain nuclear reaction cross sections differ considerably in evaluation. In addition, cross section data for all Ytterbium and many Lutetium isotopes are absent from ENDF/B-VII.1, hence JEFF 4.0 was used for \ce{^{177}Lu} production simulations. 

The results discussed in \cref{sec:Model_Comp} enable Serpent code to be applied to produce high fidelity results for the neutron flux and spectra. The results calculated in \cref{tab:VNSYields-Before-Decay} assumes that 25\% of the year, the VNS is available for continuous irradiation cycles, and presents the activity and specific activity for a variety of medical radioisotope production routes. As the VNS would be the first machine of its kind, it is important to note the capacity factor will not be known until such a device is constructed and tested, but for the purpose of this study a capacity factor must be assigned. Therefore results for capsule irradiations are presented based on what 25\% operational availability of full irradiation cycles would yield. \Cref{tab:VNSYields-After-Decay} uses literature values for individual isotopes referenced and described in the following text for processing times of established production routes to calculate the total activity and specific activity generated at the time of delivery, and simplified estimates for less-established production routes. Simplified estimate cases are treated identically with conventional production route processing times, otherwise, with one day. For short-lived diagnostic isotopes, it would be beneficial to have an on-site medical facility for their application, to minimize activity losses incurred during transport time.

Most irradiation durations were chosen to be approximately 3 half-lives, which is 90\% isotope saturation \cite{IAEA_Cyclotron2008}. Duration exceptions were made for long-lived \ce{^{60}Co}, and for byproducts including \ce{^{131}I} from fission-based \ce{^{99}Mo} production, and \ce{^{64}Cu}/\ce{^{67}Cu} which are produced together from natural Zinc.

Except for Uranium, enriched targets were idealized to 100\% enrichment. In a realistic scenario the enrichment and thus specific activities would be lower due to side reactions. Natural abundance target compositions used come from \cite{CIAAW2024}.
\\
\subsection{Specific Activity}
An important factor in determining the quality of medical isotopes produced, and whether or not the product can be used for certain applications, is specific activity. The specific activity in this work is calculated as the activity of a radioisotope of interest per unit mass of a material. One way to achieve high specific activity (HSA), is to produce the radioisotope of interest from a target composed of different element using nuclear reactions that change the proton number. This technique is called non-carrier added (NCA) production. In this study, idealized 100\% chemical separation efficiency is assumed, but in reality there are chemical impurities after separation, and very efficient separation chemistry is needed to extract the low concentration radioisotope element from the target in NCA production routes. Low specific activity (LSA) yield radioisotopes also have applications but are treated differently than HSA \cite{iaea2024}.

After chemical separation of the medical isotope element from the irradiated target, the specific activity is calculated as the target's activity of the radioisotope of interest divided by the mass of all nuclides sharing that radioisotope's element.

A variety of medical isotopes were chosen using some reactions that are unique to high energy neutrons, such as those produced in fusion reactions, as well as reactions that take place in the more conventional thermal spectrum. Some examples of typical NCA reactions performed in the high energy spectrum are $^{Z}_{N}X(n,p)_{N}^{Z-1}Y$, $^{Z}_{N}X(n,\alpha)_{N-2}^{Z-2}Y$, which often have smaller cross sections than carrier added routes such as $^{Z}_{N}X(n,2n)_{N-1}^{Z}Y$ usually is. Some reactions that typically take place in thermal regime are neutron capture (often carrier added), and fission (NCA): $^{Z}_{N}X(n,\gamma)_{N+1}^{Z}Y$; $^{Z}_{N}X(n,f)$. Carrier added routes typically require enriched targets and a high flux to achieve high specific activity, while NCA reactions require efficient chemical separation. 

\subsection{Medical Isotope Capsule Yield Analysis}
The radioisotopes selected are used for a variety of applications and have half-lives from \SI{12.7}{h} to \SI{5.27}{a}. Isotopes used for therapy predominantly emit $\alpha$, $\beta^-$, or Auger electrons, while diagnostic isotopes emit $\beta^+$ or $\gamma$ radiation and generally have shorter half-lives  \cite{iaea2024}. In this study, the therapeutic isotopes include \ce{^{131}I}, \ce{^{225}Ac}, \ce{^{177}Lu}, \ce{^{192}Ir}, \ce{^{67}Cu}, \ce{^161{Tb}}, and \ce{^{154}Sm}, and the diagnostic isotopes include \ce{^{99}Mo}, \ce{^{64}Cu}, and \ce{^{203}Pb}. \ce{^{60}Co} has a large variety of applications, for example teletherapy, sterilization, and pest control. This categorization is not comprehensive as many isotopes are used for both therapy and diagnostics.

\ce{ZrH_{1.6}} is a very efficient solid phase moderator \cite{snead2022development}. For comparison in \cref{tab:VNSYields-Before-Decay}, water was also chosen as a moderator in the case of a 19.75\% low enriched Uranium (LEU) target: \ce{UAl_2} target for \ce{^{99}Mo} production, and was found to yield $5\%$ more \ce{^{99}Mo} than in the \ce{ZrH_{1.6}} case. If \ce{ZrH_{2}} is used, a $13\%$ higher yield is achieved than water, but the Hydrogen retention is less stable \cite{snead2022development}. An unmoderated irradiation of \ce{UAl_2} was also simulated to compare fast with thermal fission yields, and it was found that even the least moderated \ce{ZrH_{1.6}} case yielded $>70\%$ more activity than unmoderated irradiation.

LEU was considered here only to allow a direct comparison with today’s Molybdenum production methods. It is recognized that the introduction of LEU in a fusion reactor, although in small amounts, would require revisiting the nuclear safety case of the facility.

\subsubsection{Molybdenum-99}
The highest \ce{^{99}Mo} yields have been found for a \ce{^{100}Mo} target utilizing the ($n,2n$) reaction, and meets the low specific activity threshold of \SI{0.05}{GBq\per\milli\gram}, needed for an LSA \ce{^{99}Mo}-\ce{^{99m}Tc} generator. Assuming idealized 100\% \ce{^{100}Mo} enrichment, a typical 6 day post-irradiation processing time \cite{ruiz1999production}, and 13 uninterrupted 7 day irradiation cycles per year, this production route could meet 74\% of the global \ce{^{99}Mo} demand in 2012 \cite{WNA2026RadioisotopesMedicine} and 85\% of the global \ce{^{99}Mo} demand in 2016 \cite{NEA2016}. Using Uranium and Ruthenium targets could respectively meet 20\%-30\% and ca. 0.1\% of global \ce{^{99}Mo} demand in 2012 and 2016 as per values calculated in \cref{tab:VNSYields-After-Decay}. The moderated \ce{^{98}Mo}($n,\gamma$)\ce{^{99}Mo} production route was also considered, which did not yield as high activities or meet the LSA threshold. A heterogeneous target involving a 50\%-50\% \ce{^{100}Mo}-\ce{^{98}Mo} target, where \ce{^{100}Mo} forms a shell around \ce{^{98}Mo}, was tested to utilize the multiplied neutrons, but the performance was still inferior to the pure \ce{^{100}Mo} target case. 

As considered by \cite{parisi2025productionhighspecificactivityradioisotopesusing}, the \ce{^{102}Ru}$(n,\alpha)$\ce{^{99}Mo} reaction was tested, and successfully yielded HSA \ce{^{99}Mo} ($>$\SI{40}{GBq\per\milli\gram} \cite{iaea2024}). The reaction cross section varies strongly depending on wether one uses JEFF 4.0 or ENDF/B-VII.1, yielding 64-\SI{87}{TBq}$\cdot$ a$^{-1}$ and 190-\SI{250}{GBq\per\milli\gram} respectively, suggesting more precise cross section measurement data would be beneficial if this production route is pursued.

\subsubsection{Iodine-131}
\ce{^{99}Mo} and \ce{^{131}I} are produced together through the fission of Uranium, and can be extracted together \cite{iaea20031340}, hence \ce{^{131}I} production using the same Uranium targets as used for \ce{^{99}Mo} . A processing time of 11 days was chosen based on the fission-based production route described in \cite{iaea20031340}. Assuming 300 days of operation per year and 4 day irradiations of six 45\% enriched Uranium plates at a time, the South African research reactor Safari-1 can produce a maximum of \SI{743}{GBq} per year \cite{iaea20031340}. Using neutron capture, a 4 day irradiation of \ce{^{nat}TeO_2} (34.48\% \ce{^{130}Te}) under a neutron flux of \SI{1e14}{ \per\centi\meter\squared\per\second} at the Budapest Research Reactor (BRR), achieved a specific activity of \SI{1,050}{GBq\per\milli\gram} Iodine. This yields \SI{74.5}{GBq} per year \cite{iaea20031340}. In both instances the potential yield of VNS using fission targets to produce \ce{^{131}I} shows potential for high specific activity and high net activity iodine production alongside Molybdenum production using an established production reaction channel.

\subsubsection{Actinium-225}
Actinium production was simulated using a \ce{RaCl_2} target (as experimentally used by \cite{APOSTOLIDIS2005383} for proton based \ce{^{225}Ac} production). At the end of irradiation, a factor of 29 times more \ce{^{227}Ac} than \ce{^{225}Ac} by mass is yielded. This is likely due to the large neutron capture cross sections of \ce{^{226}Ra}, \ce{^{225}Ac}, and \ce{^{226}Ac} \cite{Chadwick20112887}. Accumulation of \ce{^{227}Ac} is also a feature of cyclotron based production of \ce{^{225}Ac} using \ce{^{226}Ra} targets \cite{nagatsu2021cyclotron}. For this reason, activated Radium containing \ce{^{225}Ra} could be treated as the product rather than Actinium at EOI in this study, as in \cite{sasaki-2023, parisi2025productionhighspecificactivityradioisotopesusing}.  \ce{^{225}Ra} decay yields a peak \ce{^{225}Ac} value 7 days after EOI. Assuming 100\% Radium chemical separation from the irradiated target is achieved after 1 day (100\% \ce{^{225}Ac} losses before chemical separation is completed), followed by a 17 day processing time (similar to the process experimentally performed at the Japanese experimental fast reactor Joyo \cite{sasaki-2023}) the yield is 2,08\SI{0}{TBq}$\cdot$a$^{-1}$. As the greatest \ce{^{225}Ra} generator yields are when it is freshly produced, it is important that chemical separation is completed as soon as possible. If instantaneous chemical separation is assumed at EOI, an additional 4.8\% \ce{^{225}Ac} yield is calculated. 2,08\SI{0}{TBq}$\cdot$a$^{-1}$ is significantly more than current global \ce{^{225}Ac} production, which was  is estimated \SI{68}{GBq}$\cdot$a$^{-1}$ in 2017 \cite{morgenstern2018overview}, and is growing due to ongoing \ce{^{229}Th}/\ce{^{225}Ac} generator expansion at ORNL, which is planned to produce on the order of terabecquerels per year \cite{10.1021/acscentsci.0c00720}. \cite{Kratochwil1941} treated a prostate cancer patient with three \SI{6.4}{GBq} doses of \ce{^{225}Ac}-PSMA-617 and achieved complete remission. Given \SI{6.4}{GBq} dose, 2,08\SI{0}{TBq}$\cdot$a$^{-1}$ translates to 325,000 prostate cancer doses per year. If global demand increases, due to advances in \ce{^{225}Ac} clinical trials for example, irradiation in VNS has the potential to majorly contribute to the supply of \ce{^{225}Ac}.

\subsubsection{Lutetium-177}
The therapeutic isotope \ce{^{177}Lu} was produced using \ce{^{176}Lu_2O_3} and \ce{^{176}Yb_2O_3} with chemical composition also used in \cite{iaea20031340}. Both \ce{^{176}Lu} and \ce{^{176}Yb} use neutron capture, where \ce{^{176}Yb} decays into \ce{^{177}Lu} with a 1.91 hour half life \cite{IAEALiveChart}. In addition, the NCA $\ce{^{180}Hf}(n,\alpha)\ce{^{177}Yb}  \xrightarrow[T_{1/2}=1.91\ \mathrm{h}]{\beta^-}\ce{^{177}Lu}$ route possible in the fusion spectrum is considered, as discussed in \cite{parisi2025productionhighspecificactivityradioisotopesusing} using natural Hafnium targets, which has a 35.1\% abundance of \ce{^{180}Hf} \cite{CIAAW2024}. A five day decay time of the target is chosen to allow for the decay of \ce{^{176m}Lu} \cite{iaea20031340} plus additional processing time. Moderated and unmoderated target design considerations are shown. The typical specific activity yield using \ce{^{176}Lu_2O_3} targets at the end of irradiation (EOI) is \SI{2}{GBq\per\milli\gram} \cite{iaea20031340}. Half of this specific activity was achieved in VNS at EOI, but it was well exceeded in the NCA \ce{^{176}Yb_2O_3} and \ce{^{nat}Hf} target routes, yielding up to \SI{1,070}{GBq\per\milli\gram} and \SI{123}{GBq\per\milli\gram} respectively at EOI. 

It was inferred by \cite{vogel2021challenges} that the administered \ce{^{177}Lu} doses in 2021 amounted to between 74-\SI{110}{TBq}. VNS could well outproduce this in all \ce{^{176}Yb_2O_3} and \ce{^{176}Lu_2O_3} target cases, and can majorly contribute using \ce{^{nat}Hf}.

\subsubsection{Iridium-192}
\ce{^{192}Ir} production yields were calculated using \ce{^{nat}Ir} as a target using \ce{^{193}Ir}($n,2n$)\ce{^{192}Ir} and \ce{^{191}Ir}($n,\gamma$)\ce{^{192}Ir} reactions, since \ce{^{nat}Ir} is composed of 37.2\% \ce{^{191}Ir} and 62.8\% \ce{^{193}Ir} \cite{CIAAW2024}. The long lived (T$_{1/2}=73.8$ day) $\beta^-$-emitting isotope \ce{^{192}Ir} is used for brachytherapy to treat cervical, breast, and prostate cancer for example \cite{richardson-2025}. According to \cite{iaea20031340}, the Atomic Energy Department in Mumbai irradiates \ce{Na_2^{nat}IrCl_6} with a flux of 1.0-1.5$\times10^{13}$n$\cdot$cm$^{-2}$s$^{-1}$ to yield $>$\SI{0.185}{GBq\per\milli\gram} Iridium under a $30\times$ shorter irradiation exposure period than simulated in VNS. At the DHRUVA reactor in Mumbai, 25\% Iridium- 75\% Platinum wire targets are irradiated with a a flux of $1.8\times10^{14}$n$\cdot$cm$^{-2}$s$^{-1}$ to produce \SI{7.4}{GBq\per\milli\gram} specific activity \ce{^{192}Ir} and produces \SI{925}{TBq} per year \cite{sastry_kolhe_nagarja_paramr_2002}. After a cooling period of 1 day, the VNS yielded a specific activity of \SI{0.349}{GBq\per\milli\gram} and an annual production of 34,500\SI{}{TBq}, a specific activity in between the two reported irradiation scenarios, and a total annual Activity that surpasses the value reported at DHRUVA by about a factor of 37.

\subsubsection{Copper-64 and Copper-67}
\ce{^{64}Cu} and \ce{^{67}Cu} have yields have been simulated using natural Zinc targets as has also been employed for photonuclear reactions \cite{iaea2024}. These isotopes are produced together, \ce{^{64}Cu} through the \ce{^{64}Zn}$(n,p)$\ce{^{64}Cu} reaction and \ce{^{67}Cu} from \ce{^{67}Zn}$(n,p)$\ce{^{67}Cu}, where \ce{^{64}Zn} and \ce{^{67}Zn} have natural abundances of 49\% and 4\% respectively \cite{CIAAW2024}. When using Zinc targets, \ce{^{64}Cu} impurities in \ce{^{67}Cu} production can be reduced by exploiting \ce{^{64}Cu}'s shorter half life and providing a cooling time at the expense of \ce{^{67}Cu} activity, or enriched \ce{^{67}Cu} targets can be used, while only \ce{^{64}Zn} enrichment can reduce the \ce{^{67}Cu} side production of \ce{^{64}Cu}.

\ce{^{64}Cu} specific activities calculated in \cite{iaea20031340} for \ce{^{63}Cu}($n,\gamma$)\ce{^{64}Cu} at $\phi_{th}=10^{14}\text{ n}\cdot\text{cm}^{-2}\text{s}^{-1}$ yielded \SI{4}{GBq\per\milli\gram} at EOI for 1.5 days, a factor of 50 more than what was yielded for 100\% enriched \ce{^{63}Cu} in VNS. The NCA production from natural Zinc achieved much higher specific activity of \ce{^{64}Cu}, yielding 21,50\SI{0}{GBq\per\milli\gram} at the end of a 1.5 day irradiation. Given the larger cross section of the neutron capture than neutron induced proton emission reactions, the neutron capture reaction's net activity yield was more than a factor of ten larger. As \ce{^{64}Cu} is a relatively short-lived isotope ($T_{1/2}=$\SI{0.529}{d}) the delivery site should be nearby the VNS to minimize the decay losses of this isotope.

\ce{^{67}Cu} specific activities delivered using \ce{^{68}Zn}$(\gamma,p)$\ce{^{67}Cu} reactions the Low Energy Accelerator Facility (LEAF) at Argonne National Lab exceed 1,85\SI{0}{GBq\per\milli\gram} at EOI. In VNS, this exemplary specific activity was well exceeded from the 100\% enriched \ce{^{67}Zn} target case only, due to the presence of competing Copper-generating reactions in \ce{^{nat}Zn}. Additionally, LEAF produces \SI{37}{GBq} per batch \cite{iaea2024}, compared to the very large irradiation volume offered by VNS, with potential to generate about \SI{210}{TBq} per batch using \ce{^{nat}Zn} targets and 7 day irradiation cycles.

It is found that for \ce{^{nat}Zn} targets, shorter irradiations of 1.5 days with 60 cycles per year generate higher annual activities and higher specific activities for both \ce{^{64}Cu} and \ce{^{67}Cu} than 7 day irradiations with 13 cycles per year. The higher activities are due to the fact that the medical isotope itself also becomes depleted under irradiation and decay itself, with the rate of change of total isotope decreasing until it reaches saturation. The higher specific activity from shorter irradiations of \ce{^{nat}Zn} is due to an accumulation of more stable copper isotopes with respect to the \ce{^{64}Cu} or \ce{^{67}Cu} concentration within the chemically separated Copper. There may be a trade off between the cost of number of chemical separation cycles and the total activity generated which could be optimized.

\subsubsection{Terbium-161}
Four Terbium isotopes show potential for nuclear medicine application (\ce{^{149}Tb}, \ce{^{152}Tb}, \ce{^{155}Tb}, and \ce{^{161}Tb})  and theranostic pairing \cite{VanLaere2024, arman-2024}. The longest lived of which is the emerging isotope \ce{^{161}Tb}, which emits $\beta^-$, Auger electrons, and $\gamma$ radiation which would allow for therapeutic and imaging application simultaneously \cite{VanLaere2024}. To produce \ce{^{161}Tb} in the VNS, two NCA reaction pathways were considered: \ce{^{161}Dy}($n,p$)\ce{^{161}Tb} using a natural Dysprosium target (18.9\% \ce{^{161}Dy} \cite{CIAAW2024}) as also suggested for fusion reactors by \cite{parisi2025productionhighspecificactivityradioisotopesusing} and the thermal reactor neutron absorption reaction channel \ce{^{160}Gd}($n,\gamma$)\ce{^{161}Tb} using a \ce{^{160}Gd_2O_3} target as used by \cite{arman-2024}. \cite{arman-2024} yielded ca. At least \SI{2.5}{MBq} \ce{^{161}Tb} was generated from \SI{1}{mg} of target (98.2\% enriched) after 10 days of irradiation at a flux of $3\times10^{14}\text{ n}\cdot\text{cm}^{-2}\text{s}^{-1}$ at SCK CEN's BR2 reactor in Belgium. This target utilization (and implicitly, the specific activity of chemically separated terbium) is at least a factor of seven more than what is calculated for VNS after 21 days of irradiation and 5 days of cooling (\SI{0.37}{MBq\per\milli\gram} of \ce{^{161}Tb} before separation from the irradiated \ce{^{160}Gd_2O_3} target). The specific activity achieved by VNS when using \ce{^{160}Gd_2O_3} is still 50\% higher than the highest achieved specific activity of the similar half-life therapeutic isotope \ce{^{177}Lu} after processing, but yields less than 1\% of the annual activity generated for the same irradiation and processing time, and target mass. The specific activity of chemically separated Terbium from a \ce{^{nat}Dy_2O_3} target was a factor of 28 lower than that of 100\% enriched \ce{^{160}Gd}, and could likely benefit from enrichment to approach a similar specific activity level.

\subsubsection{Samarium-153}
\ce{^{153}Sm} is a theranostic isotope with $\beta^-$ and $\gamma$ emissions used for palliative bone cancer therapy, for example. The application is limited due to low specific activity of the carrier added \ce{^{152}Sm}($n,\gamma$)\ce{^{153}Sm} production route \cite{van-de-voorde-2021}. Using the \ce{^{152}Eu}($n,p$)\ce{^{152}Sm} production route \cite{parisi2025productionhighspecificactivityradioisotopesusing, sprind_ein1091}, non-carrier added \ce{^{152}Sm} can be produced. The \ce{^{nat}Eu_2O_3} (52.2\% \ce{^{152}Eu} \cite{CIAAW2024}) \cref{tab:VNSYields-Before-Decay} target yields more than a factor of 100 higher specific activity but $36\times$ less total activity from NCA production than the \ce{^{152}Sm_2O_3} target in VNS. \cite{van-de-voorde-2021} reports that the BR2 reactor yields ca. \SI{7.06}{GBq\per\milli\gram} at the end of irradiation+5 days of cooling time, which is exceeded by VNS by a factor of 8 when employing the \ce{^{nat}Eu_2O_3} target. 

\subsubsection{Lead-203}
The production of the photon emitting isotope \ce{^{203}Pb} (theranostic with $\alpha$-emitter \ce{^{212}Pb}) was also simulated via the neutron multiplying \ce{^{204}Pb}($n,2n$)\ce{^{203}Pb} reaction channel. This carrier-added route yielded a specific activity of \SI{0.032}{GBq\per\milli\gram} and total activity of \SI{3,180}{TBq} per year after an assumed one day cooling time. The specific activity of \ce{^{203}Pb} produced at TRIUMF from \ce{^{nat}Tl} targets via the NCA \ce{^{203}Tl}($p,n$)\ce{^{203}Pb} using a \SI{20}{\micro A} and \SI{12.8}{MeV} proton beam  for a 2 hour irradiation was $969.1\pm$\SI{173.9}{GBq\per\milli\gram}, four orders of magnitude higher than achieved at VNS \cite{mcneil-2023}. The advantage offered by VNS over cyclotron based production is higher net activity produced. \cite{mcneil-2023} reports after a 4 hour irradiation, a net activity with an average of $131.8\pm$\SI{4.6}{MBq} at EOI produced, compared to VNS which yields \SI{292}{TBq} at EOI after one 6 day irradiation cycle, six orders of magnitude more total activity generated per unit time, but dispersed over a far larger target volume. 

\refstepcounter{footnote}
\label{isotope_info_foot}
\footnotetext[1]{The half-life values are taken from the IAEA Livechart \cite{IAEALiveChart} }

\refstepcounter{footnote}
\label{SA_foot}
\footnotetext[2]{The specific activity as presented in \cref{tab:VNSYields-Before-Decay} and \cref{tab:VNSYields-After-Decay} assume ideal 100\% efficient chemical separation of the radioisotope element from other elements in the target. Isotope separation is not considered}

\refstepcounter{footnote}
\label{H2O_foot}
\footnotetext[3]{\ce{H_2O} moderated}

\refstepcounter{footnote}
\label{ZrH2_foot}
\footnotetext[4]{\ce{ZrH_2} moderated}

\refstepcounter{footnote}
\label{JEFF4_foot}
\footnotetext[5]{Denoted calculations use JEFF 4.0 nuclear data library instead of ENDF/B-VII.1}

\begin{tabularx}{\textwidth}{XXXXXXXX}
\caption{Calculated VNS Medical Isotope Annual Yields at 25\% Irradiation Availability, without considering post-irradiation decay losses. Unspecified moderators correspond to \ce{ZrH_{1.6}}, - corresponds to no moderator. Uranium targets presented in the table are enriched to 19.75\%.}
\label{tab:VNSYields-Before-Decay} \\
    \hline\hline
    Medical Isotope Produced ($T_{1/2}$ [d]\textsuperscript{\hyperref[isotope_info_foot]{a}}) & Target Material &  Moderator Volume per Capsule [cm$^3$] & Irradiation Cycle Length [Days]  & Irradiation Cycles per Year (25\% Operation) & Total Activity \newline [TBq$\cdot$a$^{-1}$] & Specific Activity in Element [GBq$\cdot$mg$^{-1}$]\textsuperscript{\hyperref[SA_foot]{b}} \\
    \hline
     \ce{^{99}Mo} & \ce{UAl_2} & 2.72 & 7 & 13 & 23,900 & 2,560 \\
      (2.75) & \ce{UAl_2} & 2.72\textsuperscript{\hyperref[H2O_foot]{c}} & 7 & 13 & 25,200 & 2,560 \\
      & \ce{UAl_2} & 2.72\textsuperscript{\hyperref[ZrH2_foot]{d}} & 7 & 13 & 27,000 & 2,550 \\
      & \ce{UAl_2} & - & 7 & 13 & 14,200 & 2,560 \\
      & \ce{^{100}Mo} & - & 7 & 13 & 77,000 & 0.059 \\
      & \ce{^{98}Mo}   & 2.9 & 7 & 13 & 26,000 & 0.023 \\
      & \ce{^{100}Mo-^{98}Mo} & - & 7 & 13 & 51,800 & 0.041  \\
      & \ce{^{nat}Ru} & - & 7 & 13 & 87 & 250 \\
      & \ce{^{nat}Ru}\textsuperscript{\hyperref[JEFF4_foot]{e}} & - & 7 & 13 & 64 & 190 \\ \hline
	\ce{^{131}I} (8.03) & \ce{UAl_2} & 2.72 & 7 & 13 & 6,270 & 2,190\\
    & \ce{UAl_2} & - & 7 & 13 & 3,810 & 2,160\\\hline
    \ce{^{225}Ra} (14.9) & \ce{^{226}RaCl_2} & - & 30 & 4 & 1,220 & 0.0161 \\ \hline
    \ce{^{177}Lu} & \ce{^{176}Yb_2O_3}\textsuperscript{\hyperref[JEFF4_foot]{e}}& - & 21 & 4 & 7,130 & 778 \\
     (6.64) & \ce{^{176}Yb_2O_3}\textsuperscript{\hyperref[JEFF4_foot]{e}}& 2.7 & 21 & 4 & 10,100 & 1,070 \\
      & \ce{^{176}Lu_2O_3}\textsuperscript{\hyperref[JEFF4_foot]{e}}& - & 21 & 4 & 306,000 & 1 \\
      & \ce{^{176}Lu_2O_3}\textsuperscript{\hyperref[JEFF4_foot]{e}}& 2.5 & 21 & 4 & 443,000 & 1 \\
      & \ce{^{nat}Hf}\textsuperscript{\hyperref[JEFF4_foot]{e}}& - & 21 & 4 & 18.4 & 123 \\ \hline
    \ce{^{192}Ir} (73.8) & \ce{^{nat}Ir} & - & 210 & 1 & 34,900 & 0.352 \\ \hline
    \ce{^{64}Cu}  & \ce{^{nat}Zn} & - & 1.5 & 60 & 41,900 & 21,500 \\
     (0.529) & \ce{^{nat}Zn} & - & 7 & 13 & 10,600 & 5,990 \\
      & \ce{^{63}Cu} & 2.73 & 1.5 & 60 & 497,000 & 0.084 \\ \hline
    \ce{^{67}Cu} & \ce{^{nat}Zn} & - & 1.5 & 60 & 386 & 198 \\
     (2.58) & \ce{^{nat}Zn} & - & 7 & 13 & 215 & 121 \\
      & \ce{^{67}Zn} & - & 7 & 13 & 3,550 & 28,000 \\ \hline 
    \ce{^{161}Tb} & \ce{^{nat}Dy_2O_3} & - & 21 & 4 & 12.5 & 57.6 \\
    (6.89)  & \ce{^{160}Gd_2O_3} & 2.94 & 21 & 4 & 73 & 1,400 \\ \hline
    \ce{^{153}Sm} & \ce{^{152}Sm_2O_3} & 2.94 & 6 & 15 & 3,390 & 4.3 \\
     (1.93) & \ce{^{nat}Eu_2O_3} & - & 6 & 15 & 92 & 584 \\ \hline
	\ce{^{203}Pb} (2.16) & \ce{^{204}Pb} & - & 6 & 15 & 4,380 & 0.04 \\ \hline
    \hline
\end{tabularx}

\begin{tabularx}{\textwidth}{XXXXXXXX}
\caption{Calculated VNS Medical Isotope Annual Yields after decay losses from processing times, at 25\% Irradiation Availability. Unspecified moderators correspond to \ce{ZrH_{1.6}}, - corresponds to no moderator. \Cref{tab:VNSYields-Before-Decay} contains information regarding irradiation cycles/year. Uranium targets presented in the table are enriched to 19.75\%.}
\label{tab:VNSYields-After-Decay} \\
    \hline\hline
    Medical Isotope Produced ($T_{1/2}$ [d]\textsuperscript{\hyperref[isotope_info_foot]{a}}) & Target Material &  Moderator Volume per Capsule [cm$^3$] & Irradiation Cycle Length [Days] & Processing Time Assumed [Days] & Total Activity [TBq$\cdot$a$^{-1}$ & Specific Activity in Element [GBq$\cdot$mg$^{-1}$]\textsuperscript{\hyperref[SA_foot]{b}} \\
    \hline
      \ce{^{99}Mo} & \ce{UAl_2} & 2.72 & 7 & 6 & 5,260 & 600 \\
      (2.75) & \ce{UAl_2} & 2.72\textsuperscript{\hyperref[H2O_foot]{c}} & 7  & 6 & 5,560 & 600 \\
      & \ce{UAl_2} & 2.72\textsuperscript{\hyperref[ZrH2_foot]{d}} & 7 & 6 & 5,950 & 600 \\
      & \ce{UAl_2} & - & 7 & 6 & 3,120 & 600 \\
 & \ce{^{100}Mo} & - & 7 & 6\textsuperscript{\hyperref[proctime_foot]{f}}& 17,000 & 0.013             \\
 & \ce{^{98}Mo} & 2.9 & 7 &  6\textsuperscript{\hyperref[proctime_foot]{f}} & 5,700 & 0.005           \\
& \ce{^{100}Mo-^{98}Mo} & - & 7 & 6\textsuperscript{\hyperref[proctime_foot]{f}} & 11,400 & 0.009 \\
 & \ce{^{nat}Ru} & - & 7 & 6\textsuperscript{\hyperref[proctime_foot]{f}} & 19 & 62                  \\
 & \ce{^{nat}Ru}\textsuperscript{\hyperref[JEFF4_foot]{e}} & - & 7 & 6\textsuperscript{\hyperref[proctime_foot]{f}} & 14 & 46                  \\\hline
\ce{^{131}I} (8.03) & \ce{UAl_2} & 2.72 & 7 & 11 & 2,530 & 2,290 \\
\ce{^{131}I} & \ce{UAl_2} & - & 7 & 11 & 1,540 & 2,130 \\\hline
\ce{^{225}Ac} (9.92) & \ce{^{226}RaCl_2} & - & 30 & 18\textsuperscript{\hyperref[Ac_foot]{g}} & 2,080 & 2,150           \\\hline
\ce{^{177}Lu} & \ce{^{176}Yb_2O_3}\textsuperscript{\hyperref[JEFF4_foot]{e}} & - & 21 & 5 & 4,290 & 425          \\
(6.64) & \ce{^{176}Yb_2O_3}\textsuperscript{\hyperref[JEFF4_foot]{e}} & 2.7 & 21 & 5 & 6,200 & 613        \\
 & \ce{^{176}Lu_2O_3}\textsuperscript{\hyperref[JEFF4_foot]{e}} & - & 21 & 5 & 182,000 & 0.52       \\
 & \ce{^{176}Lu_2O_3}\textsuperscript{\hyperref[JEFF4_foot]{e}} & 2.5 & 21 & 5 & 263,000 & 0.754    \\
 & \ce{^{nat}Hf}\textsuperscript{\hyperref[JEFF4_foot]{e}} & - & 21 & 5\textsuperscript{\hyperref[proctime_foot]{f}} & 11 & 53.6                \\\hline
\ce{^{192}Ir} (73.8) & \ce{^{nat}Ir} & - & 210 & 1\textsuperscript{\hyperref[proctime_foot]{f}} & 34,500 & 0.349            \\\hline
\ce{^{64}Cu} & \ce{^{nat}Zn} & - & 1.5 & 1\textsuperscript{\hyperref[proctime_foot]{f}} & 11,300 & 6,480              \\
(0.529) & \ce{^{nat}Zn} & - & 7 & 1\textsuperscript{\hyperref[proctime_foot]{f}} & 2,847 & 1,660               \\
 & \ce{^{63}Cu} & 2.73 & 1.5 & 1\textsuperscript{\hyperref[proctime_foot]{f}} & 134,000 & 0.0225         \\\hline
\ce{^{67}Cu} & \ce{^{nat}Zn} & - & 1.5 & 1\textsuperscript{\hyperref[proctime_foot]{f}} & 295 & 169               \\
(2.58) & \ce{^{nat}Zn} & - & 7 & 1\textsuperscript{\hyperref[proctime_foot]{f}} & 164 & 95.3              \\
 & \ce{^{67}Zn} & - & 7 & 1\textsuperscript{\hyperref[proctime_foot]{f}} & 2,710 & 27,900         \\\hline
\ce{^{161}Tb} & \ce{^{nat}Dy_2O_3} & - & 21 & 5 & 7.55 & 33.2      \\
(6.89) & \ce{^{160}Gd_2O_3} & 2.94 & 21 & 5 & 44.0 & 932  \\\hline
\ce{^{153}Sm} & \ce{^{152}Sm_2O_3} & 2.94 & 6 & 5 & 562 & 0.389 \\
(1.93) & \ce{^{nat}Eu_2O_3} & - & 6 & 5 & 15.3 & 56.7     \\\hline
\ce{^{203}Pb} (2.16) & \ce{^{204}Pb} & - & 6 & 1\textsuperscript{\hyperref[proctime_foot]{f}} & 3,180 & 0.032 \\\hline
    \hline
\end{tabularx}

\refstepcounter{footnote}
\phantomsection\label{proctime_foot}
\footnotetext[6]{Denoted processing times are assumed and not necessarily well-established}
\refstepcounter{footnote}
\phantomsection\label{Ac_foot}
\footnotetext[7]{Irradiated Radium cools one day after EOI, and then is chemically separated to 100\% Radium (incurring loss of \ce{^{225}Ac} generated in target until that point). The chemically pure activated Radium target then generates \ce{^{225}Ac}, with maximum activity 18 days after EOI.}

\subsection{Capsule Irradiation Nuclear Heating}
It is important to quantify the heat produced in capsules for design and safety considerations. Two cases were considered, \ce{^{100}Mo} and \ce{UAl_2} targets as representative of the range of energy release since \ce{^{100}Mo}($n,2n$)\ce{^{99}Mo} is an endothermic reaction, with an \SI{8.5}{MeV} threshold energy \cite{Chadwick20112887} while Uranium fission releases approximately \SI{200}{MeV}. The heating power was calculated using neutron and photon induced heating detectors (ENDF neutron MT-4 and photon MT-12 reaction numbers), with results presented in \cref{tab:caps_heat}. \ce{UAl_2} generates \SI{17.8}{W} per fresh capsule exposed to flux. This heat must be extracted by the water flowing around the capsules in the pipes. If the whole irradiation facility is filled with such fission target capsules, the collective nuclear power generated contributes a substantial \SI{1.42}{\mega\watt} to the VNS system, while \ce{^{100}Mo} capsules would contribute \SI{0.27}{\mega\watt}. \Cref{tab:caps_heat} contains calculations of heat transfer across the capsule shell and from the capsule to surrounding water. These calculations assume a steady-state, so capsules entering the inlet to the irradiation zone should always have the same initial temperature, although they may change temperature as they move along the system. The conductive heat transfer across the \SI{0.5}{\milli\meter} thick shell is found to have a difference of $<\SI{0.5}{\kelvin}$. Using a very conservative convective heat transfer coefficient $h=\SI{20}{\watt\per\meter\squared\per\kelvin}$, temperature differences of \SI{6.88}{\kelvin} and \SI{175}{\kelvin} are calculated for \ce{^{100}Mo} and \ce{UAl_2} targets respectively. %

For target materials not containing fissile materials the heat generation is sufficiently low to not heat up the capsules by more than \SI{10}{\kelvin}. In contrast, capsules containing LEU were predicted to generate approximately 20 times more heat. Further studies are needed to assess the expected temperature increase of LEU-containing capsules and it may be necessary to lower the assumed  \SI{1.1}{g} of LEU per capsule to limit the heat produced.

\begin{table}
\centering
\begin{tabularx}{0.5\textwidth}{|p{0.2\textwidth}|XX|}%
\caption{Spherical capsule heat for \ce{^{100}Mo} and \ce{H_2O}-moderated \ce{UAl_2}. The symbols represent as follows: Target heat power density $\dot{q}_{target}$, Non-target capsule heat power $\dot{Q}_{struc}$, $\dot{Q}_{caps}$, Capsule surface area A\textsubscript{ext}, capsule diameter D\textsubscript{ext}, capsule wall thickness  t, convective heat transfer coefficient $h$, temperature difference between capsule shell and water $\Delta$T\textsubscript{conv}, capsule wall thermal conductivity $\kappa$, temperature difference across capsule shell $\Delta$T\textsubscript{shell}.}\label{tab:caps_heat}\\
    \hline\rule{0pt}{10pt}
    Target & \ce{^{100}Mo} & \ce{UAl_2} \\\hline
    \rule{0pt}{10pt}$\dot{q}_{target}$ [\SI{}{\watt\per\gram}] & 0.011 & 12.7 \\
    $\dot{Q}_{struc}$ [\SI{}{\watt}] & 0.689 & 0.63 \\ %
    m\textsubscript{target} [\SI{}{\gram}] & 1.1 &  \\
    $\dot{Q}_{caps}$  [\SI{}{\watt}] & 0.701 & 17.8 \\
    A\textsubscript{ext} [\SI{}{\milli\meter\squared}] & 1,018 & \\
    D\textsubscript{ext} [\SI{}{\milli\meter}] & 18 & \\
    t [mm] & 0.5 & \\
    $h$ [\SI{}{\watt\per\meter\squared\per\kelvin}] & 100 & \\
    $\kappa$ [\SI{}{\watt\per\meter\per\kelvin}] & 20.0 &   \\
    $\Delta$T\textsubscript{conv} [\SI{}{\kelvin}] & 6.88 & 175 \\
    $\Delta$T\textsubscript{shell} [\SI{}{\kelvin}] & 0.017  & 0.44 \\\hline
\end{tabularx}
\end{table}

\subsection{Cobalt Plate Irradiation Yield Analysis}\label{sec:Co_Yield}
\begin{tabularx}{\textwidth}{XXXXXX}
\caption{2-Year irradiation yields of out-board blanket Cobalt plate designs at end of irradiation}
\label{tab:cobalt}
\\\hline\hline
    Cobalt Plate Thickness [cm] & \ce{ZrH_{1.6}} Plate Thickness [cm] & Cobalt Loading [kg] & Activity at EOI [TBq] & Specific Activity [GBq$\cdot$mg$^{-1}$]\\\hline
    1 & - & 810 & 1,900 & 0.023 \\\hline
    10 & - & 8,600 & 7,500 & 0.0087 \\\hline
    10 & 3 & 9,000 & 15,000 & 0.017 \\\hline
    0.1 & 3 & 84 & 2,100 & 0.25 \\\hline\hline
\end{tabularx}
\begin{tabularx}{\textwidth}{XXXXXXX}
\caption{Further cobalt plate optimization: 3-year irradiation yields of out-board blanket design scenarios at end of irradiation}
\label{tab:cobalt-opti}
\\\hline\hline
    Cobalt Plate Thickness [cm] & \ce{H_2O} Thickness Before Plate [cm] & \ce{H_2O} Thickness Behind Plate [cm] & Cobalt Loading [kg] & Activity at EOI [TBq] & Specific Activity [GBq$\cdot$mg$^{-1}$]\\\hline
    0.1 & 3 & - &  83.6 & 30,000 & 0.36 \\\hline
    0.1 & 6 & - &  86.6 & 37,000 & 0.42 \\
\hline
    0.1 & 3 & 3 & 83.6 & 86,000 & 1.04 \\
\hline
    0.1 & 3 & 30 & 83.6 & 100,000 & 1.2 \\\hline
    0.1 & 16.5 & 16.5 &  98.8 & 54,000 & 0.54 \\\hline
    0.1 & 30 & 3 &  97.1 & 21,000 & 0.21 \\\hline
    3 & 15 & 15 &  2,965 & 120,000 & 0.042 \\\hline
    1 & 3 & 29 & 843 & 190,000 & 0.22 \\
\hline\hline
\end{tabularx}
The lower region of the outboard blanket may be available for additional irradiation volume, as the higher flux upper part of the blanket is used for blanket testing and various ports. \ce{^{60}Co} activity yields and specific activities for four initial Cobalt plate designs are shown in \cref{tab:cobalt}. A major advantage of \ce{^{60}Co} production is that natural Cobalt is mono-isotopic, consisting of \ce{^{59}Co} \cite{CIAAW2024}, which has a large thermal neutron capture cross section $\sigma_{n,\gamma}$(\SI{25}{meV})$=$\SI{37}{b} \cite{Chadwick20112887}. A continuous irradiation of two full power years is considered to estimate VNS potential for \ce{^{60}Co} production. Because of \ce{^{60}Co}'s long 5.3 year half-life \cite{IAEALiveChart}, VNS plant outages are expected to have a weaker effect on the activity yields than the production of shorter-lived radioisotopes which is discussed in further detail in \cref{sec:dwell}. The specific activity did not reach reach the HSA threshold of approximately $>$\SI{3.7}{GBq\per\milli\gram} \cite{ArchiveMarketResearch2026_HSACo60} even for the highest specific activity design scenario in which a \SI{1}{mm} thin Cobalt plate was irradiated behind a \SI{3}{cm} thick \ce{ZrH_{1.6}} moderator. In nuclear reactors where HSA \ce{^{60}Co} is produced, a longer irradiation time is taken, for example CANDU and RMBK reactors take 1-3 and 5 years respectively \cite{MYLVAGANAM1990602, WNA2026RadioisotopesMedicine} in addition to utilizing a thermal neutron flux. The VNS \SI{1}{mm} thin Cobalt plate case was designed to mitigate the effect of self shielding, in which the flux facing part of the plate blocks absorption from the back of the plate, which indeed managed to increase the specific activity by a factor of 150 for a 1,000 fold thinning of the plate. Further optimization is performed by increasing moderation for a more fully thermalized flux, and more finely comparing the dependency of specific activity on Cobalt plate thickness to achieve a higher net activity yield in \cref{tab:cobalt-opti}. According to \cite{MYLVAGANAM1990602}, CANDU reactors produce between 2.22-\SI{9.25}{GBq\per\milli\gram} and approximately 140,00\SI{0}{TBq} per year. In comparison, the highest specific activity VNS case yielded \SI{1.2}{GBq\per\milli\gram} and 100,00\SI{0}{TBq} after a three year irradiation period. Further optimization must be carried out for the VNS to reach \ce{^{60}Co} production comparable with CANDU reactors.

It is important to consider that the irradiation along the plate is not perfectly homogeneous. Top and bottom \SI{2}{\centi\meter} flux voxels were compared with the middle segment of the \SI{80}{\centi\meter} tall cobalt plate described by row four of \cref{tab:cobalt-opti} and it was found that the top segment experienced a flux of $[7.6\pm0.02]\times10^{13}$ n$\cdot $cm$^{-2}$s$^{-1}$ (114\% of middle of the plate), while the bottom segment experienced a flux of $[6.2\pm0.01]\times10^{13}$n$\cdot $cm$^{-2}$s$^{-1}$ (93\% of middle segment). This as expected from the poloidial flux distribution seen in \cref{fig:xz_nflux}, there is lower flux in the bottom part of the plate. It is foreseeable that such an irradiated plate could be axially chopped up such that the higher specific activity segment may be separated from the lower specific activity region. In the case described by row four in \cref{tab:cobalt-opti} for example, when comparing the specific activity of the top and bottom of the Cobalt plate, the bottom is found to have a specific activity of \SI{2.4}{GBq\per\milli\gram}, a factor of two larger than the average and top segment of the plate which coincidentally are equal. In the case described by row seven in \cref{tab:cobalt-opti} on the other hand, when comparing the specific activity of the top and bottom of the Cobalt plate, the top is found to have a specific activity of \SI{0.54}{GBq\per\milli\gram}, a factor of two larger than the bottom segment of the plate, and 16\% higher than the plate average. Specific activity strongly depends on the Cobalt plate geometry, and many factors play a role in the optimization.

\subsection{VNS Outage Dwell Time Case Studies}\label{sec:dwell}
\begin{figure}[htb]
    \centering
    \begin{subfigure}[t]{=0.4\textwidth}
    	\centering
        \includegraphics[width=0.9\textwidth]{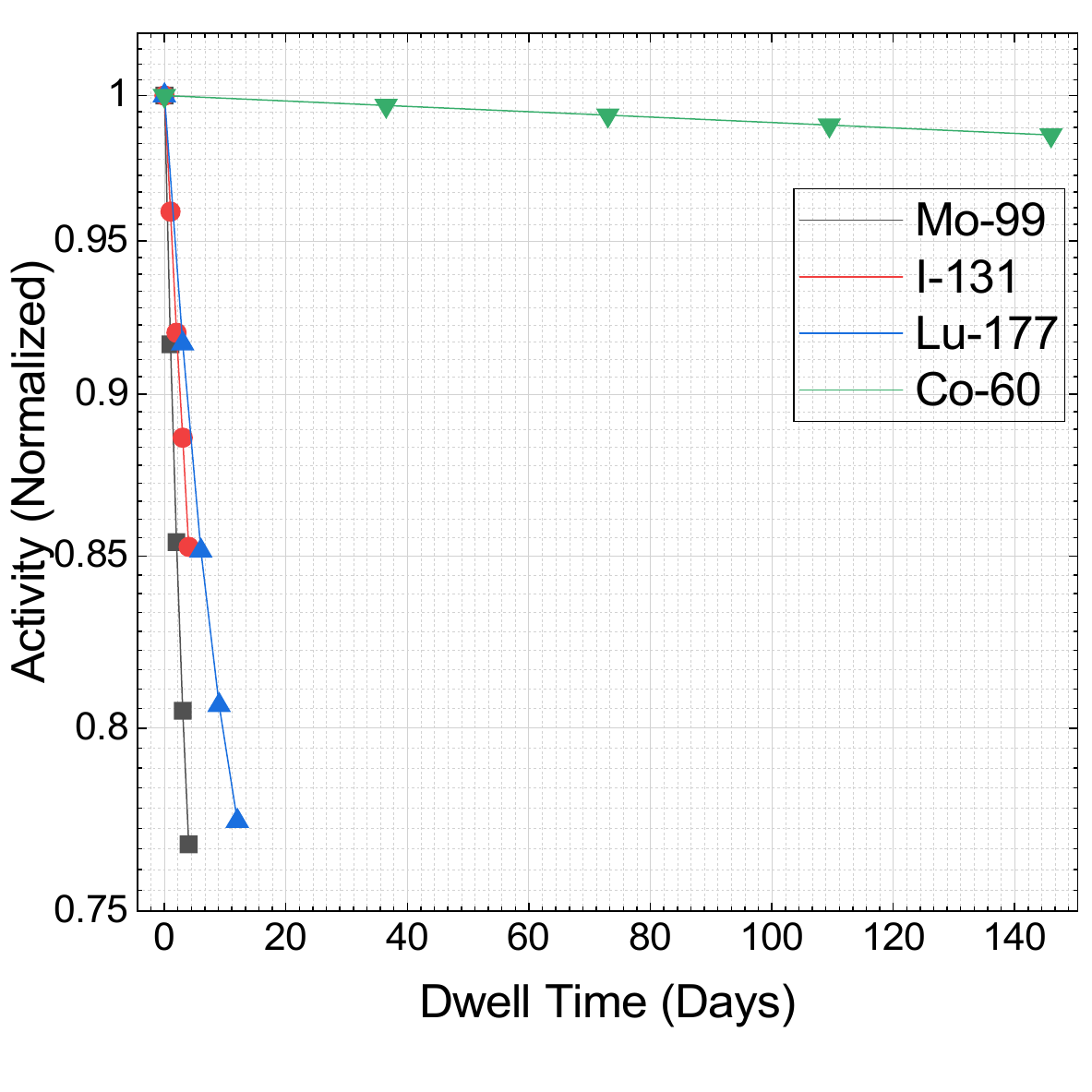}
        \caption{Activity vs. Dwell Time (Days)}
        \label{fig:dwell_Day}
    \end{subfigure}\quad
    \begin{subfigure}[t]{=0.4\textwidth}
    	\centering
        \includegraphics[width=0.9\textwidth]{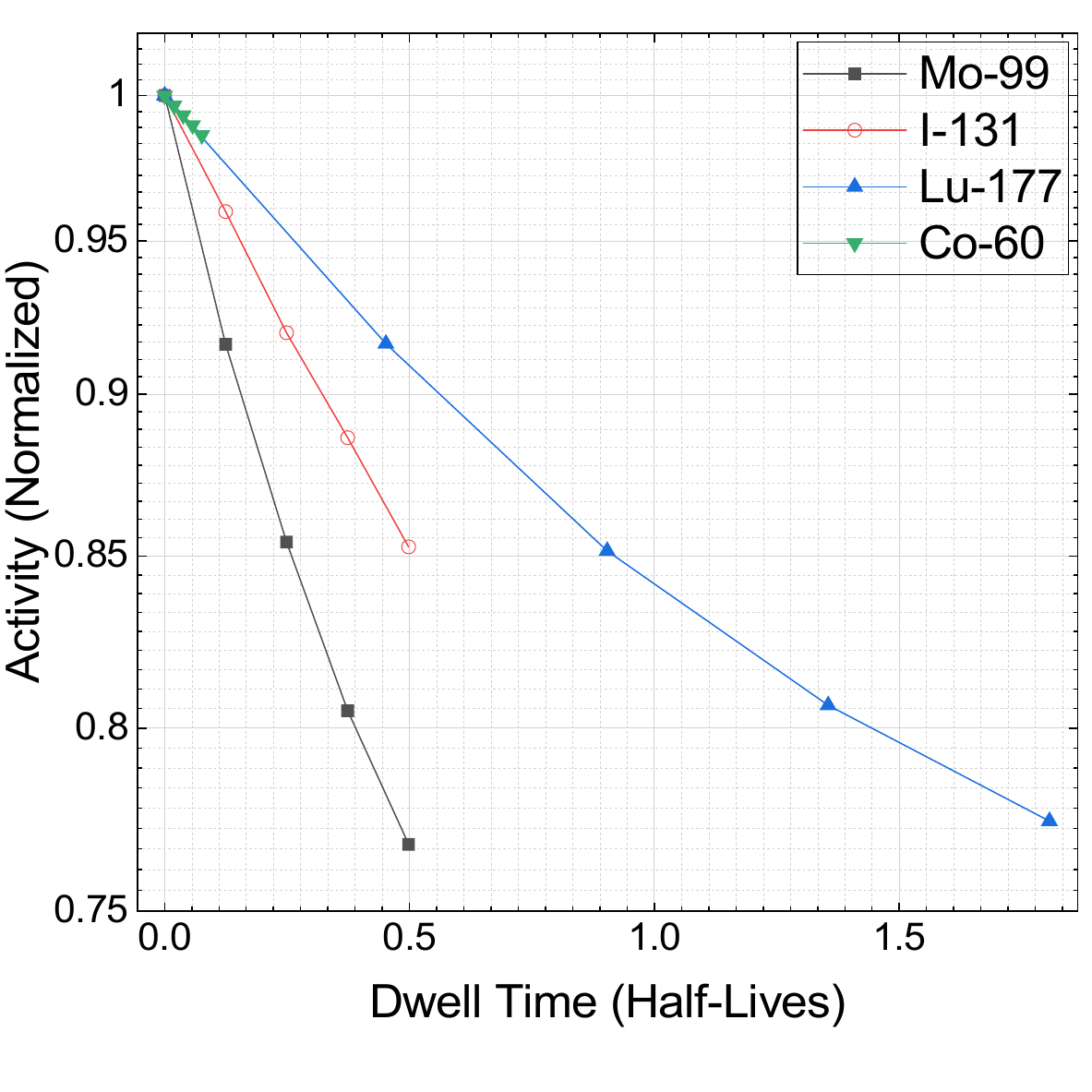}
        \caption{Activity vs. Dwell Time (Half-lives)}
        \label{fig:dwell_HL}
    \end{subfigure}
    \label{fig:dwell-time}
    \caption{Activity generated normalized to continuous operation scenario (no dwell time) as a function of VNS dwell time for \ce{^{99}Mo}, \ce{^{131}I}, \ce{^{177}Lu}, \ce{^{60}Co} (start of dwell times at 4, 4, 12, and 365 days after irradiation start respectively.).}
    \label{fig:activity-vs-dwelltimedays}
\end{figure}
As the VNS would be a first of a kind machine, it is very likely to experience outages, at least in the initial operation phases. Therefore, an analysis on the outage effects on medical isotope yields should be studied. As there are many different possibilities for how plant a plant outage could manifest, a few simple cases were considered with the same neutron fluence at the EOI which involved 100\% operation dropping instantaneously to 0\% power, and then after a dwell time, returning to 100\% as a function of reactor outage duration. \ce{^{99}Mo}, \ce{^{131}I} ($T_{1/2}=$\SI{2.75}{\day} \cite{IAEALiveChart}), ($T_{1/2}=$\SI{8.03}{\day} \cite{IAEALiveChart}) and \ce{^{177}Lu} ($T_{1/2}=$\SI{6.64}{\day} \cite{IAEALiveChart}), and \ce{^{60}Co} ($T_{1/2}=$\SI{1925}{\day} \cite{IAEALiveChart}) were selected for this case study shown in \cref{fig:activity-vs-dwelltimedays}. It can be seen that relatively long-lived \ce{^{60}Co} can withstand much longer outages than \ce{^{99}Mo}, \ce{^{131}I} and \ce{^{177}Lu} from \cref{fig:dwell_Day}, while calculation results shown in \cref{fig:dwell_HL} normalize VNS dwell time to the half-life of respective isotopes and shows that \ce{^{177}Lu} can be regenerated per unit half-life. After 146 days 98.6\% of the \ce{^{60}Co} activity yield remains, while after shorter outage periods, eg. 4 days for \ce{^{99}Mo} and \ce{^{131}I}, 76.8\% and 85.3\% of the activity remains, incurring a far greater loss. \ce{^{177}Lu} losses were considered for up to 12 days of dwell time, where in the case of 12 days 77.4\% of the activity yield remained. After 3 days of dwell time and restart, \ce{^{99}Mo}, \ce{^{131}I} and \ce{^{177}Lu} lose 20\%, 12.4\%, 19.9\% respectively. For a system such as the VNS, in addition to selecting isotopes with cross sections advantageous for the spectrum, candidate isotopes should be robust to plant outages.

\section{CONCLUSIONS}
Good agreement has been found for neutron and photon transport between OpenMC 0.15.2 and Serpent 2.2.2 models of the VNS vacuum vessel, with photon and neutron flux relative difference varying within the statistical uncertainties. The ($n,2n$) reaction had the highest difference for a reaction rate, with OpenMC predicting $[5.4\pm1.4]\%$ lower than Serpent and photon while the photon flux relative difference was smaller than standard error. For default tracking modes of coupled neutron-photon simulation, Serpent was between 1.6 and $1.7\times$ faster than OpenMC, although OpenMC is $1.6\times$ faster in neutron only transport mode. 

For medical isotope production, different particle source types have their own advantages. For example, medical cyclotrons are compact and mature, and can be placed directly in hospitals to reliably deliver medical isotopes including short lived imaging isotopes which must be quickly transported to the patients. Research reactors have the ability to reliably deliver high neutron flux and there exists a variety of thermal reactions with high cross sections. Fission reactors are a mature technology, and function as central production facilities, which are especially well suited for isotopes with half-lives long enough to for transportation and processing, such as \ce{^{99}Mo} ($T_{1/2}=2.75$) and \ce{^{177}Lu} ($T_{1/2}=6.64$) for example. A tokamak fusion neutron source would be a new technology, and is likely to face unsteady operation until the technology matures, but has the advantages of large irradiation volume availability, a high energy neutron spectrum which can perform non-carrier added medical isotope production routes, and could potentially offer a synergy with breeding blanket technology via neutron multiplication reactions. 

A variety of production routes for eleven radioisotopes were simulated in the VNS Tokamak, showcasing it as a machine with radioisotope production potential, provided stable operation and efficient target-isotope separation chemistry. Production of \ce{^{99}Mo}, \ce{^{131}I}, \ce{^{225}Ac}, \ce{^{177}Lu}, \ce{^{192}Ir}, \ce{^{64}Cu}, \ce{^{67}Cu}, \ce{^{161}Tb}, \ce{^{153}Sm}, and \ce{^{203}Pb} were simulated in a capsule irradiation facility under the assumption of full length irradiation cycles at 25\% VNS annual availability, while \ce{^{60}Co} production was simulated for plate based irradiation under full VNS irradiation capacity. Under the assumption of a 100\% enriched \ce{^{100}Mo} target, it was calculated that a full capsule loading of VNS could produce 17,000 six-day TBq \ce{^{99}Mo} annually at a specific activity of \SI{0.013}{GBq\per\milli\gram}, that is the equivalent of 85\% of the global \ce{^{99}Mo} demand in 2019 using the \ce{^{100}Mo}($n,2n$)\ce{^{99}Mo} reaction channel and low specific activity \ce{^{99}Mo}-\ce{^{99m}Tc} generators. Furthermore, non-carrier added reaction channels requiring high energy neutrons available in fusion reactors such as \ce{^{102}Ru}($n,\alpha$)\ce{^{99}Mo} can produce high specific activity \ce{^{99}Mo} if efficient chemical separation can be performed, in the case of 100\% chemical separation 14-19 six-day TBq of \ce{^{99}Mo} annually at a specific activity of 46-\SI{62}{GBq\per\milli\gram}. An activity yield difference of 26.4\% lower was found in the JEFF 4.0 than the ENDF/B-VII.1 nuclear data library for this \ce{^{102}Ru}($n,\alpha$)\ce{^{99}Mo} reaction, highlighting the importance of accurate nuclear data in the fusion neutron spectrum to alleviate discrepancies, which could be supported by a high intensity fusion neutron source such as that which could be provided by the VNS.

Using a water-moderated 19.75\% enriched \ce{UAl_2} target, a yield of 5,560 six-day TBq \ce{^{99}Mo} annually at a high specific activity of \SI{600}{GBq\per\milli\gram}, that is approximately 32\% of the 2016 global demand \cite{NEA2016}. 

Simulated \ce{^{225}Ac} production yielded sufficient Activity to accommodate 325,000 \SI{6.4}{GBq} doses per year using \ce{^{226}RaCl_2} targets. 

After 5 days of processing, 6,20\SI{0}{TBq} per year of high specific activity (\SI{613}{GBq\per\milli\gram}) \ce{^{177}Lu} can be generated using moderated targets by virtue of the high irradiation volume allotted by VNS using neutron capture on a moderated \ce{^{176}Yb_2O_3} target, which far surpasses the 74-\SI{110}{TBq} administered globally in 2021 inferred by \cite{vogel2021challenges}. Alternatively \ce{^{nat}Hf} can be used as a target for the high threshold energy $\ce{^{180}Hf}(n,\alpha)\ce{^{177}Yb}  \xrightarrow[T_{1/2}=1.91\ \mathrm{h}]{\beta^-}\ce{^{177}Lu}$ reaction to capitalize on the high energy neutrons provided by VNS. \ce{^{nat}Hf} irradiation in VNS yields \SI{11}{TBq} per year (10-15\% of the aforementioned inferred 2021 globally administered activity) at a specific activity \SI{53.6}{GBq\per\milli\gram}.

Calculation of \ce{^{192}Ir} production at VNS yielded an annual production a factor of 37 greater than produced at the DHRUVA reactor in India, and at a specific activity a factor of $21\times$ lower, but still higher than the specific activity yielded for another procedure reported by the Atomic Energy Department in Mumbai, indicating potential for use for VNS-produced Iridium. Additionally, the long 73.8 day half-life makes \ce{^{192}Ir} an isotope more robust to VNS dwell time than shorter lived isotopes such as \ce{^{99}Mo}.

\ce{Zn} targets can be used to produce radioisotopes \ce{^{64}Cu}, and \ce{^{67}Cu} via ($n,p$) reactions. After 100\% efficient separation from target with an assumed 1 day from end of irradiation to delivery, a specific activity of 21,50\SI{0}{GBq\per\milli\gram} (a factor of 50 higher \ce{^{64}Cu}) is yielded than using a moderated \ce{^{63}Cu} target in VNS. As \ce{^{64}Cu} is a short lived ($T_{1/2}=$\SI{0.529}{d}) isotope, the delivery site should be near, if not at the VNS site for the deployment of this isotope. The VNS exceeded \ce{^{67}Cu} specific activity of Argonne National Lab's LEAF facility by 190\% when using an idealized 100\% enriched \ce{^{67}Zn} target, but not for \ce{^{nat}Zn} which yielded 11\% of the LEAF value. This suggests up to one order of magnitude specific activity can be gained through \ce{Zn} target enrichment at VNS. Furthermore the VNS  \ce{^{67}Zn} target yields more than three orders of magnitude more activity per batch than LEAF, while the VNS \ce{^{nat}Zn} yielded up to $250\times$ more activity than LEAF, suggesting abundant \ce{^{67}Cu} production potential in both schema.

VNS production of \ce{^{161}Tb} yielded a specific activity of \SI{932}{GBq\per\milli\gram} when using \ce{^{160}Gd_2O_3} targets, a factor of at least 7 lower than what can be produced by BR2, but still 50\% higher than the similar half-life and also therapeutic isotope \ce{^{177}Lu} in the best VNS case. The total activity generated by VNS for the best \ce{^{161}Tb} case was still a factor of 28 lower than the best VNS \ce{^{177}Lu} case.

Simulated production of \ce{^{152}Sm} in VNS exceeded the specific activity produced by BR2 by a factor of 8 via the usage of \ce{^{nat}Eu_2O_3} targets for the non-carrier added \ce{^{152}Eu}($n,p$)\ce{^{152}Sm} reaction channel. This indicates the VNS could be a strong candidate for HSA \ce{^{152}Sm} production, provided the operation conditions and chemical separation assumptions made in this study.

\ce{^{203}Pb} carrier-added production calculated in VNS yielded four orders of magnitude lower specific activity but six orders of magnitude total activity per irradiation time than NCA cyclotron based production of \ce{^{203}Pb} at TRIUMF. This suggests \ce{^{203}Pb} as a weak candidate for production in VNS, unless application is found at the specific activity generated.

Nuclear heating was evaluated for LEU and \ce{^{100}Mo} target material capsules and it was found that \ce{^{100}Mo} capsules heat the surrounding water less than \SI{10}{\kelvin} while LEU requires additional cooling or less than \SI{1.1}{g} of target material.

In addition to the capsule irradiation facility modeled in the in-board blanket, a Cobalt plate was modeled in the available volume in the lower part of the out-board blanket. The plate was optimized by adjusting the plate thickness, position of the plate within the moderator, moderator thickness, choice of moderator, and irradiation time. The maximum specific activity achieved was \SI{1.2}{GBq\per\milli\gram} with a total of 100,00\SI{0}{GBq} after three years of irradiation. This is not far from the approximate \SI{3.7}{GBq\per\milli\gram} HSA \ce{^{60}Co} threshold, indicating potential for VNS as a supplier of \ce{^{60}Co}, however higher specific activity and net annual activity can be produced by CANDU reactors.

VNS outages were simulated for \ce{^{99}Mo}, \ce{^{131}I}, \ce{^{177}Lu}, and \ce{^{60}Co} production under various dwell time scenarios and found that it is possible to restart irradiation and salvage medical isotope production at the cost of activity loss, with longer lived isotopes being more robust to plant outages. An optimization study was carried out for the Cobalt plate and it was found that moderation and plate thickness optimization could improve the irradiated plate's specific activity and activity by more than one order of magnitude. It is likely that the capsule irradiation facility can also be further neutronically optimized for high activity yields. 

\section*{ACKNOWLEDGEMENTS}
This work has been carried out within the framework of the EUROfusion Consortium, funded by the European Union via the Euratom Research and Training Programme (Grant Agreement No 101052200 — EUROfusion). Views and opinions expressed are however those of the author(s) only and do not necessarily reflect those of the European Union or the European Commission. Neither the European Union nor the European Commission can be held responsible for them. The authors gratefully acknowledge the computational and data resources provided by the Leibniz Supercomputing Centre (www.lrz.de). The authors would like to thank colleagues Gabriele Burgio, Lilianna Quintero-Sombrano, Daniel Bonete-Wiese, Christof Löwenhag, and Michele Lungaroni for their support during the preparation of this work.

\printbibliography
\end{document}